\documentclass[preprint,12pt]{elsarticle}

\usepackage{latexsym}
\usepackage{amssymb}

\usepackage{amssymb,amsthm}
\usepackage{amsmath}
\usepackage{amsfonts}

\usepackage{ulem}

\usepackage{algorithm}
\usepackage{algpseudocode}

\usepackage{tikz}
\usetikzlibrary{decorations.pathreplacing,snakes,mindmap}
\usetikzlibrary{arrows, calc, decorations.pathmorphing}

\newtheorem{conj}{Conjecture}[section]

\begin{document}

\begin{frontmatter}

\title{$k$-path graphs: experiments and conjectures about algebraic connectivity and $\alpha$-index}

\author{Rafael L. de Paula$^{1}$} \author{Claudia M. Justel$^{2}$}
\author{Carla S. Oliveira$^{3}$} \author{Milena S. Caruba$^{4}$} 

\address{$^{1}$ Departamento de Engenharia de Produção, CEFET Rio de Janeiro, Brazil. 
}
\address{$^{2}$ PGSC - Departamento de Engenharia de Computa\c c\~ao, Instituto Militar de Engenharia, Rio de Janeiro, Brazil.
}
\address{$^{3}$ Departamento de Matemática\\Escola Nacional de Ciências Estatisticas, IBGE, Rio de Janeiro, Brazil.
}
\address{$^{4}$ Departamento de Engenharia de Computa\c c\~ao, Instituto Militar de Engenharia, Rio de Janeiro, Brazil.
}

\begin{abstract}
This work presents conjectures about eigenvalues of matrices associated with $k$-path graphs, the algebraic connectivity, defined as the second smallest eigenvalue of the Laplacian matrix, and the $\alpha$-index, as the largest eigenvalue of the $A_{\alpha}$-matrix. For this purpose, a process based on [Discrete Applied Mathematics 164 (2014) 297-303] is presented to generate lists of $k$-path graphs containing all non-isomorphic 2-paths, 3-paths, and 4-paths of order $n$, for $6 \leq n \leq 26, 8 \leq n \leq 19$, and $10 \leq n \leq 18$, respectively. Using these lists, exhaustive searches for extremal graphs of fixed order for the mentioned eigenvalues were performed.
Based on the empirical results, conjectures are suggested about the structure of extremal $k$-path graphs for these eigenvalues. 
\end{abstract}



\end{frontmatter}

\section{Introduction}

Spectral Graph Theory (SGT) studies the relationship between the structural properties of graphs and the eigenvalues of matrices associated with them. One line of research in SGT is the determination of extremal graphs by analyzing their eigenvalues. From this perspective, we can cite several works related to some particular classes of graphs, such as paths, trees, unicyclic, bicyclic, and planar graphs, as can be seen in \cite{Tait2017},  \cite{ABREU2014429}, \cite{yu:15}, and \cite{Yu:19}.

Among the Laplacian matrix's eigenvalues, the second smallest one stands out because it has a fundamental property related to the connectivity of the graph and is called algebraic connectivity. This eigenvalue is useful in several applications, for instance, the connectivity control of mobile robot networks, in \cite{Ikemoto2020}; the analysis of complex directed networks with scale-free properties, in \cite{imae2018}, and image segmentation, in \cite{Grady2006}. Regarding $A_{\alpha}$-matrix, it was proposed by \cite{Nikiforov:17} as a generalization of the signless Laplacian matrix, $Q(G)$. The signless Laplacian matrix $Q(G)$, proposed by \cite{Cvetkovic:05}, is considered by Nikiforov to be a unique matrix in several aspects, since it is the sum of the diagonal degree and adjacency matrices of a graph. Its study revealed similarities and differences between these two matrices. Currently, \cite{Nikiforov:17} seminal article has more than 240 citations in the Scopus database, giving rise to a line of research on linear combinations of matrices associated with graphs. Several papers about eigenvalues of $A_{\alpha}$-matrix, in particular the largest eigenvalue, are highlighted in the bibliography.

$k$-Path graphs are an important subfamily of $k$-trees, which, in turn, are chordal graphs.
$k$-Trees are a generalization of trees; a single vertex is a tree; adding an adjacent vertex to a single vertex of a tree with $n$ vertices generates a new tree with $n + 1$ vertices. The definition of $k$-trees given in \cite{Rose1974} is a generalization of the previous definition: a $k$-tree of $k$ vertices is a complete graph of $k$ vertices. Let $T$ be a $k$-tree with $n$ vertices, then a $k$-tree $T^{\prime}$ with $n+1$ vertices is obtained from $T$ by connecting a new vertex to each vertex of an existing clique of size $k$. Several problems that are $NP$-complete for graphs in general admit polynomial algorithms to solve them, when restricted to the class of chordal graphs, in particular $k$-trees (and in some cases linear ones). For this reason, chordal graphs and their subclasses are of considerable interest in the graph structure research line.

According to \cite{Brinkmann:13}, using lists of graphs can help as a source of ideas to formulate conjectures or possible counterexamples in the study of Graph Theory. There are several web pages containing lists of graphs (e.g., the lists by Brendan McKay\footnote{http://cs.anu.edu.au/$\sim$bdm/data/}, Gordon Royle\footnote{http://mapleta.maths.uwa.edu.au/$\sim$gordon/}, Markus Meringer\footnote{http://www.mathe2.unibayreuth.de/markus/reggraphs.html}, Frank Ruskey\footnote{http://www.theory.cs.uvic.ca/$\sim$cos/}, and Ted Spence\footnote{http://www.maths.gla.ac.uk/$\sim$es/}. The book \textit{Atlas of Graphs} (\cite{Read:98}) contains lists of drawings of various graphs, including their adjacency lists. More recently, the latest version of the graph database \textit{House of Graphs} (\cite{Coolsaet:23}) provides a metadirectory containing lists of all non-isomorphic graphs up to a given order for various graph classes. The directory allows searching for graphs in the available lists.

The objective of this work is to produce lists of $k$-path graphs of bounded order, for $k \in \{2, 3, 4\}$ in $g6$ format. To our knowledge, there are no graph lists in $g6$ format available for $k$-path graphs. By using these lists, we look for information about the structure of fixed order $k$-path graphs that attain the maximum and minimum values of the algebraic connectivity, and $\alpha$-index. In a previous paper (\cite{Justel:23}), the particular case of $2$-path graphs of bounded order was analyzed by using lists of graphs in $g6$ format, and conjectures about the maximum value of the algebraic connectivity  was stated.  
Unlike this new article, in that previous publication, the maximum value of algebraic connectivity was also analyzed for maximal outerplanar graphs of bounded order, by using lists of graphs in $g6$ format. 

This paper is organized as follows. Section~\ref{sec:conceitos} presents basic concepts on Graph Theory and Spectral Graph Theory. In Section~\ref{sec:metodologia}, information about the construction of lists of $k$-path graphs for fixed and limited $k$ and $n$ is introduced. Section~\ref{sec:resulteexperimentos} presents the results obtained for the experiments executed with the lists of $k$-path graphs. In Sections~\ref{sec:metodologia} and~\ref{sec:resulteexperimentos}, the contributions of this work are given. Final remarks are presented in  Section~\ref{sec:concideracoesFinais}.

\section{Basic Concepts}\label{sec:conceitos}

Let $G=(V,E)$ be an undirected graph with $|V|=n$ vertices, called the {\bf graph order}, and $|E|=m$ edges. The neighborhood of $v \in V$ is denoted $Adj(v) = \{w \in V \mid \{v,w\} \in E\}$. The degree of $v$, denoted $d_v$, is the cardinality of its neighborhood, that is, $d_v = |Adj(v)|$. If $w \in Adj(v)$, we say that $w$ is adjacent to $v$.
Given $S \subset V$, the subgraph of $G$ induced by $S$ is denoted $G[S]$. If $G[S]$ is a complete subgraph, then $S$ is called a {\bf clique}. A clique of size $k$ is denoted a $k$-clique. A clique is \textbf{maximal} if it is not a proper subset (different from $\emptyset$ and the set itself) of any other clique. We denote by $K_n$, $P_n$, and $C_n$, respectively, the complete, path, and cycle graphs of order $n$. A vertex $v \in V$ is {\bf simplicial} in $G$ when $Adj(v)$ is a clique in $G$;  $v$ is {\bf universal} when $d_v = n - 1$. The \textbf{distance} between two vertices $u$ and $v$ is the length of the shortest path between them, denoted by $d(u,v)$. If there is no path between two vertices, $d(u,v) = \infty$.
The \textbf{vertex connectivity} of a graph, denoted by $\kappa(G)$, is the smallest number of vertices that, when removed, make the graph disconnected.
\indent A {\bf $k$-coloring} of a graph $G$ consists of assigning to each $v \in V$ a color, from a set of $k$ colors, in such a way that adjacent vertices receive different colors; the graph is then said to be \textbf{$k$-colorable}. The \textbf{chromatic number} of a graph is the smallest $k$ such that the graph is $k$-colorable and is denoted by $\chi(G)$. If $G = K_n$, its chromatic number is $\chi(K_n) = n$.

\indent If $G_1=(V_1,E_1)$ and $G_2=(V_2,E_2)$ are vertex-disjoint graphs ($V_1 \cap V_2 = \emptyset$), their graph sum is $G_1 + G_2 = (V_1 \cup V_2, E_1 \cup E_2)$. The {\bf join} operation between two graphs, $G_1 \vee G_2$, is the graph obtained from $G_1 + G_2$ by adding new edges between each vertex of $G_1$ and all vertices of $G_2$.
Given a graph $G$ and a positive integer $d$, the {\bf $d$-power} of $G = (V, E)$ is the graph $G^d = (V, E')$  such that two vertices are adjacent in $G^d$ if and only if the distance between them in $G$ is at most $d$.
In this work, we will frequently use the graphs $K_{k-1} \vee P_{n-k+1}$ and $P_n^k$, the latter called $k$-ribbon in \cite{Markenzon:14}.
For instance, the graph $K_1 \vee P_{n-1}$ is called a \textbf{fan} of order $n$, and the graph $P_n^{2}$ is called a \textbf{2-ribbon} of order $n$.

\indent A graph $G = (V, E)$ is {\bf chordal} if every cycle of size greater than or equal to 4 admits a chord (that is, an edge whose endpoints are non-consecutive vertices in the cycle).

A subset $S \subset V$ is a \textbf{vertex separator} for a pair of non-adjacent vertices $u$ and $v$ (a $uv$-separator) if removing $S$ from the graph separates $u$ and $v$ into distinct connected components. A \textbf{minimal $uv$-separator} is a $uv$-separator that does not contain another $uv$-separator as a proper subset. When the vertex pair is not mentioned, $S$ is called a \textbf{minimal vertex separator}.
A \textbf{clique tree} of $G$ is a tree $T$ whose vertices are the maximal cliques of $G$ such that for every pair of maximal cliques $Q, Q'$, the path between them in the tree $T$ contains the intersection of these two cliques. According to \cite{Blair93}, clique trees of chordal graphs allow represent the graph compactly and efficiently. Given a chordal graph, there exists a set of clique trees associated with this graph. It can be proved that verifying that $S \subset V$ is a minimal vertex separator is equivalent to verifying $S=Q \cap Q'$ for some pair of maximal cliques $Q, Q'$, that represent vertices in a clique tree of a chordal graph. Furthermore, the set of minimal vertex separators of the chordal graph is represented in this way in any clique tree.
Other definitions and results on graphs and chordal graphs can be found in \cite{DIESTEL:01}, \cite{Blair93}, and \cite{GOLUMBIC:80}.

\indent The {\bf Laplacian matrix} of a graph $G=(V, E)$ of order $n$, is defined as $L(G) = D(G) - A(G)$, 
where $D(G) = diag(d_{v_1}, \dots, d_{v_n})$ is the diagonal degree matrix of the graph's vertices and $A(G)$ is its adjacency matrix. $L(G)$ is symmetric, 
so it has $n$ real eigenvalues, and it is also positive semidefinite, so all its eigenvalues are nonnegative. Furthermore, the matrix $L(G)$ is singular, and therefore the smallest eigenvalue is equal to zero. 
The eigenvalues of $L(G)$ are denoted by  $\mu_1(G)\geq \dots \geq \mu_{n-1}(G) \geq \mu_n(G)  = 0$. 
In \cite{Fiedler:73} it is proved that a graph is connected if and only if $\mu_{n-1}(G) > 0$. This eigenvalue is called {\bf algebraic connectivity} and is denoted by $a(G)$. Fiedler also proved that the vertex connectivity is an upper bound for the algebraic connectivity of non-complete graphs.

The \textbf{signless Laplacian matrix} of a graph $G$, denoted by $Q(G)$, is obtained from the sum of the diagonal matrix and the adjacency matrix, $Q(G) = D(G) + A(G)$. The characteristic polynomial of this matrix $Q$ 
is denoted by $p_Q(\lambda)$,  and its eigenvalues (also known as $Q$-eigenvalues) are denoted by $q_1(G) \geq q_2(G) \geq \dots \geq q_n(G)$, being  $q_1(G)$ known as the \textbf{index} of $Q(G)$ or \textbf{spectral radius} of the signless Laplacian matrix. 

The \textbf{$A_{\alpha}(G)$ matrix} is the result of a convex combination between the diagonal matrix degree and the adjacency matrix, considering any $0 \leq \alpha \leq 1$: $A_{\alpha}(G)  = \alpha D(G) + (1-\alpha) A(G)$. The eigenvalues of $A_{\alpha}(G) $
are  denoted by $\lambda_1(A_{\alpha}) \geq \lambda_2(A_{\alpha}) \geq ...\geq \lambda_n(A_{\alpha})$.  
The spectral radius of this matrix ($\lambda_1(A_{\alpha})$) is called {$\alpha$-index}. Some important relations to be highlighted are:
\[
A(G) = A_0(G),\,\, D(G) = A_1(G) \,\, \text{e} \,\, Q(G) = 2A_{1/2}(G).
\]

\indent Other definitions and results on matrices and eigenvalues can be found in \cite{DEABREU:07}, \cite{Cvetkovic:05}, and \cite{Nikiforov:17}.

\subsection{{$k$-path graphs}}

In the same article that presents the inductive definition of $k$-trees, \cite{Rose1974}, the following characterization of these graphs is presented. A connected graph that has a $k$-clique and has no $k+2$-clique, and for which all minimal separators induce a complete graph $K_k$, is a $k$-tree.

According to \cite{Proskurowski1999}, $k$-path graphs can be defined recursively as follows:  let $n \geq k$ and $k \geq 2$, every complete graph with $k$ vertices is a $k$-path graph.
If $G = (V, E)$ is a $k$-path graph, and $Q \subset V$ is a $k$-clique containing at least one simplicial vertex, then for any vertex $v \notin V$, the augmented graph $G' = (V \cup \{v\}, E \cup \{ \{v,w\} \mid w \in Q\}$ also is a $k$-path graph.

For $k$-trees, which are chordal graphs, the number of edges and the chromatic number are known (\cite{Rose1974}, \cite{GOLUMBIC:80}).
Since $k$-path graphs verify the definition of $k$-tree, can be concluded that $k$-path graphs have $kn - \frac{k(k+1)}{2}$ edges and chromatic number $k+1$.
In this work, we consider $k$-paths for $k\geq2$ and $n \geq k+1$.

In \cite{Markenzon:06}, it is proved that maximal outerplanar graphs are 2-trees for which the clique chosen to augment the graph was not previously chosen. 2-path graphs verify this property; therefore, 2-path graphs are maximal outerplanar. In the same paper, it is proved that planar 3-trees are exactly chordal and maximal planar graphs.
Since 3-paths are chordal graphs, their number of edges is $3n-6$, and they can be drawn in the plane without edge crossing, we can conclude that they are maximal planar graphs, in particular, planar. Observe that for $k \geq 4$, $k$-paths are non-planar graphs.

\subsection{Related works}

In this section, we present results from the literature on spectral parameters for $k$-paths and $k$-trees.

Let $\mathbf{T}_{n}^k$ be the set of all $k$-trees of order $n$. Spectral properties of $k$-trees are presented in \cite{ZHANG2015}. Specifically, the $k$-tree graph of fixed order with the maximum index of the signless Laplacian matrix, $q_1(G)$, is characterized, and an optimal upper bound for $q_1(G)$ in $\mathbf{T}_{n}^k$ is determined. Furthermore,  upper bounds for $q_1(G)$ (and for related expressions such as $q_1(G)$ + $k$, $q_1(G)$ - $k$, $q_1(G)$ . $k$, $q_1(G)/k$) are determined for graphs $G$ belonging to the set in $\bigcup_{k=1}^{n-1}$ $\mathbf{T}_{n}^k$.

In \cite{yu:15}, the authors defined $\mathbf{G}_n$ as the set of maximal outerplanar graphs with $n$ vertices and no inner triangles (which are exactly the 2-path graphs). Then they proved that,   
the fixed order $n$ graphs in $\mathbf{G}_n$  with $q_1(G)$ (the largest eigenvalue of the index of the signless Laplacian matrix) maximum and minimum are, respectively, $K_1 \vee P_{n-1}$ and $P_n^2$.

In \cite{Lin:18}, the spectrum of the matrix $A_{\alpha}$ for some classes of graphs is considered; in particular, they determine the spectrum of the complete split graph, denoted $K_s \vee (n-s)K_1$, which is a particular case of $k$-tree. 

In \cite{Yu:19}, the authors prove that the unique outerplanar graph of order $n$ with a maximum value of $\alpha$-index is $K_1 \vee P_{n-1}$. Furthermore, they prove that, if $\alpha \in [0.486, 0.5]$, the unique planar graph of order $n$ with maximum $\alpha$-index is $K_2 \vee P_{n-2}$.

In \cite{OLIVEIRA:21} the authors analyze the nullity (the multiplicity of zero as an eigenvalue of $A(G)$) of two families of $k$-trees, the $(k+1)$-line graphs of $k$-trees and the free block-connected graphs $K_{1,k+2}$, and the simple clique $k$-trees, which are the $k$-trees whose $(k+1)$-line graphs are trees. For each pair of integers $p \geq 2$ and $k \geq 2$, a $k$-tree is presented such that the nullity of its $(k+1)$-row graph is equal to $p$. Furthermore, the authors establish an upper bound for the nullity of the $(k+1)-$row graphs of $k$-simple clique trees with $n$ vertices. Also, the paper demonstrates the existence of $k$-simple clique trees whose $(k+1)-$row graphs have nullity between 1 and $n-(k+2)-2\lceil\frac{n}{k+1}\rceil$.

In \cite{Justel:23}
 were created lists of graphs containing maximal outerplanar graphs (mops) with fixed order, ranging from 6 to 11, and containing 2-path graphs with fixed order, ranging from 6 to 24. For mops, lists of planar graphs of House of Graphs were used, and the maximal outerplanar graphs obtained by filtering those lists. For 2-path graphs sequences of integer numbers between 1, 2 and 3 representing non-isomorphic graphs of this class as in \cite{PEREIRA:14} were used. The maximum value of the algebraic connectivity of mops and 2-path graphs of limited order were analyzed.

\subsection{Generating non-isomorphic $k$-path graphs}

In \cite{PEREIRA:14} is presented a procedure that allows, given $k \geq 2$, to generate and count all non-isomorphic $k$-path graphs of fixed order. To this end, a color sequence obtained from a $(k+1)$-coloring of the $k$-path graph, induced by its clique tree, is used. Regarding that, according to \cite{PEREIRA:14}, in the case of $k$-path graphs, the clique tree is a path.

The \textbf{core sequence} is given by $n-k$ ordered maximal cliques corresponding to the clique tree of the $k$-path graph. Denote the core sequence by $\mathcal{F}(G) = \langle F_1, F_2, \dots , F_{n-k} \rangle$,  where $|F_i| = k+1, \forall \, 1 \leq i \leq n-k$. It is worth noting that for a $k$-path graph, there are two core sequences, because given a core sequence $\mathcal{F}(G)$, its reverse sequence, $\mathcal{F'}(G) = \langle F_{n-k}, \dots, F_1\rangle $, is also a core sequence of $G$.

The intersection of two consecutive cliques in $\mathcal{F}(G)$ is a $k$-clique. Denote these intersections by $B_i = F_{i+1} \cap F_i$, for $i=1,\dots,n-k-1$. On the other hand, the difference of the sets corresponding to consecutive cliques, $F_{i+1} \backslash F_i $, contains a single vertex.

Let $G$ be a $k$-path graph. Since $G$ is $(k + 1)$-colorable, there exists a $k + 1$-coloring of $G$. The authors \cite{PEREIRA:14} define the \textbf{color sequence} $C = \langle c_1, c_2, \dots, c_{n - k - 1} \rangle$, where $c_i$ is the color corresponding, by the $(k+1)$-coloring, to the vertex $v_i$ in the sequence $\langle v_1, v_2, \dots, v_{n - k - 1} \rangle$ with $v_i$ being the only vertex in $F_{i + 1} \backslash F_i$. If $n = k + 1$, the color sequence is empty.

Two new concepts are introduced by \cite{PEREIRA:14} in order to characterize $k$-path graphs of order $n$ by sequences of length $n - k + 1$. Given a sequence of integers of length $j$, $C = \langle a_1, a_2, \dots, a_j\rangle$, $C$ is said to be \textbf{normalized} if $a_1 = 1$ and $a_i \leq 1 + \max(a_1, \dots, a_{i-1})$ for $2 \leq i \leq j$. Furthermore, if sequence $C$ satisfies $a_i \neq a_{i+1}$ for $1 \leq i \leq j - 1$ is said to be \textbf{restricted}. By using these definitions, the authors proved the two Lemmas below.

{\bf Lemma 2} (Lemma 4 \cite{PEREIRA:14}) Let $\mathcal{F}(G) = \langle F_1, F_2, \dots , F_{n-k}\rangle$ be a core sequence of a $k$-path graph of order $n$. There is a unique restricted normalized color sequence $C = \langle c_1, c_2, \dots , c_{n-k-1}\rangle$ of $\mathcal{F}(G)$ with at most $k + 1$ colors.

{\bf Lemma 3} (Lemma 5 \cite{PEREIRA:14}) Let $C = \langle a_1, a_2, \dots , a_{n-k-1}\rangle$ be a restricted normalized sequence with at most $k+1$ distinct integers. $C$ is a color sequence of a unique $k$-path graph of order $n$.

In \cite{PEREIRA:14}, the total number of $k$-path graphs of order $n$ and given $k$, denoted  $T(n,k)$, is computed. Table~\ref{tab:Tnk} shows 
the corresponding values of $T(n,k)$, for $1 \leq k \leq 4$.
In \cite{B22}, the same number of unlabeled $k$-paths
of order $n$ is also obtained by using a different approach.

\begin{table}[h] 
\centering 
\begin{small}
\caption{ $T(n, k)$ for $1 \leq k \leq 4$ from Table 3 \cite{PEREIRA:14}.} 
\begin{tabular}{|c|c|c|}
\hline
& even $n$ & odd $n$ \\ \hline
$k=1$ & $1$ & $1$ \\ \hline
$k=2$ & $2^{n-6} + 2^{\frac{n-6}{2}}$ & $2^{n-6} + 2^{\frac{n-7}{2}}$ \\ \hline
$k=3$ & $\frac{3^{n-6} + 2 \cdot 3^{\frac{n-6}{2}} + 1}{4}$ & $\frac{3^{n-6} + 4 \cdot 3^{\frac{n-7}{2}} + 1}{4}$ \\ \hline
$k=4$ & $\frac{4^{n-8} + 4 \cdot 2^{n-8} + 1}{3}$ & $\frac{4^{n-8} + 7 \cdot 2^{n-9} + 1}{3}$ \\ \hline
\end{tabular}
\label{tab:Tnk}
\end{small}
\end{table}

In the next section, we will present the algorithm proposed by \cite{PEREIRA:14} (based on Lemmas 2 and 4) and how it will be used to produce the lists of $k$-path graphs with fixed and limited $k$ and $n$.

\section{Methodology}\label{sec:metodologia}

\begin{figure}[htb]
\begin{center}
\includegraphics[width=1\textwidth]{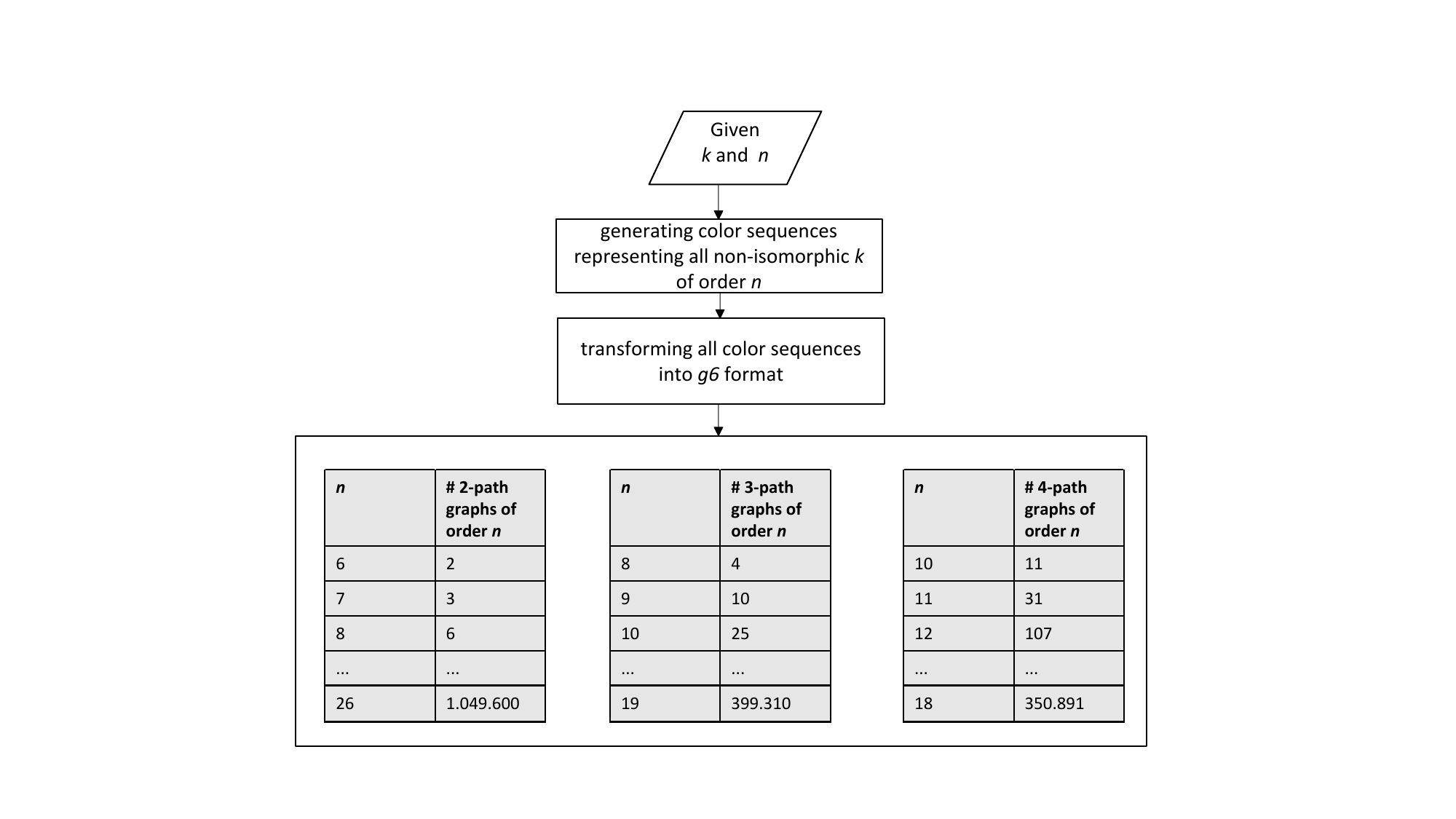}
 \caption{Construction's phases for $k$-path lists $L_{k,n}$ with $k \in \{2, 3, 4\}$, fixed and bounded $n$}
 \label{processo_generico1}
\end{center}
\end{figure}

\begin{figure}[htb]
\centering 
 \includegraphics[width=1\textwidth]{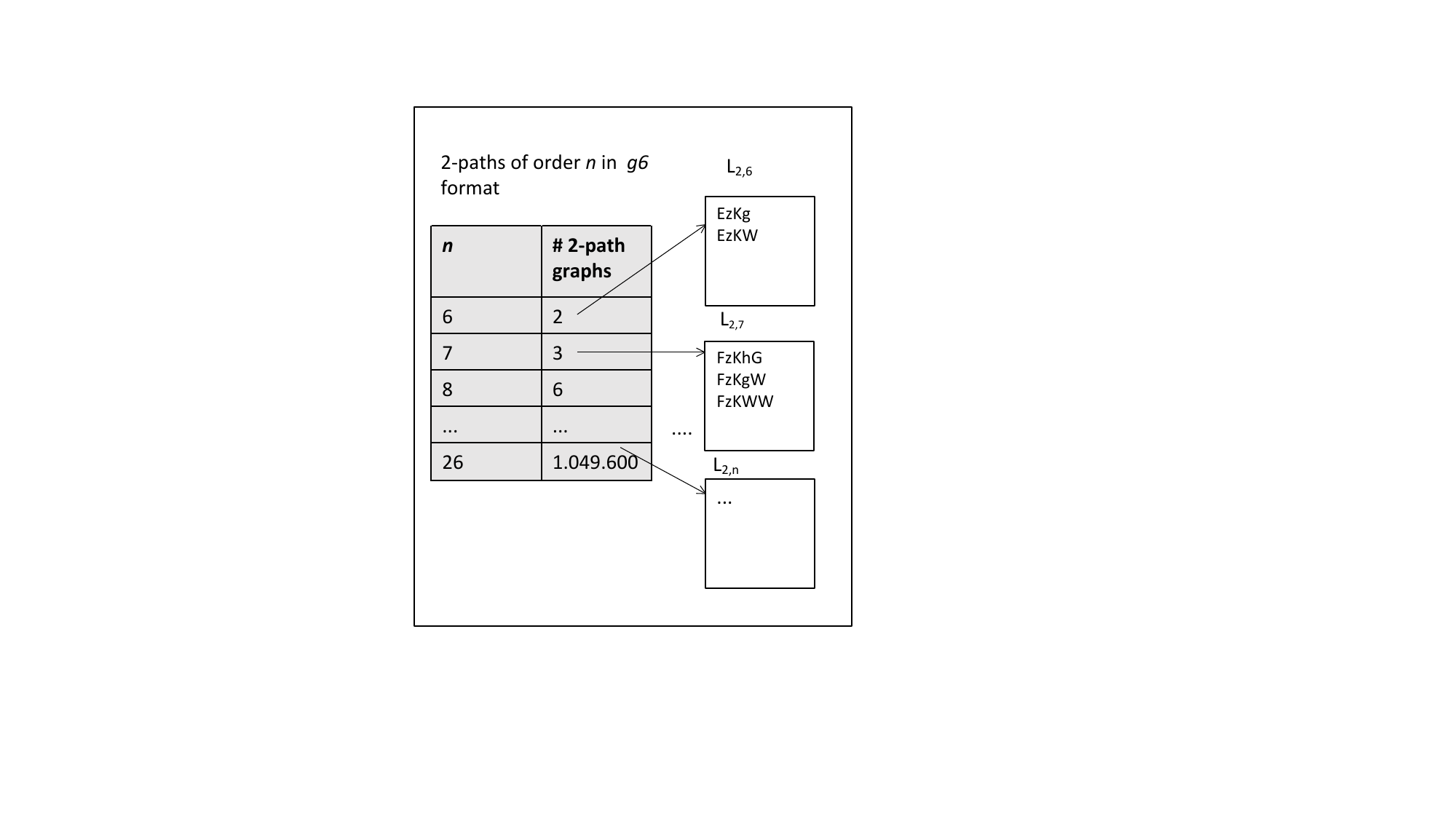}
 \caption{Details 2-path graphs lists $L_{2,n}$; formats: $txt$ for lists and $g6$ for graphs}
 \label{processo_generico2}
\end{figure}

This section presents the 3 phases for constructing the lists of $k$-path graphs generated in this work. Figure~\ref {processo_generico1} illustrates the construction. Given valid $k$ and $n$:

\noindent phase i) Generation of color sequences representing all non-isomorphic $k$-path graphs of order $n$.  The generation of $k$-path graphs represented by color sequences, as proposed in \cite{PEREIRA:14}, is used. Subsection~\ref{detalhes_etapa_a} describes the details of the algorithm implemented for this purpose.\newline
phase ii) For each graph, transformation of the color sequence generated in phase i) to $graph6$ ($g6$) format.  Subsection~\ref{detalhes_etapa_b} presents this transformation.
\newline
phase iii) For fixed $k$  and $n$, storage of the lists of $k$-path graphs with limited number of vertices $n$ in $txt$ files. Each list is denoted $L_{k,n}$, and each graph is stored in $g6$ format (see Figure~\ref{processo_generico2} and Subsection~\ref{detalhes_etapa_c}). 
The values of $k$  considered were $k = 2, 3, 4$. The bounds on $n$ will be discussed in Section~\ref{sec:resulteexperimentos}.
\newline
All algorithms were implemented in a computer with a 3.7 GHz AMD Ryzen 5 5600X processor and 16 GB of RAM, using the Python (v3.12.7) programming language.  \texttt{NetworkX} library (v3.3) was used to generate, manipulate, and analyze the graphs. 
The graph images were produced with \texttt{TikZ} library, through the \texttt{tikzpicture} environment, in \LaTeX.

\pagebreak

\subsection{Color Sequence Generation - phase i}\label{detalhes_etapa_a}

In Section 2.3, theoretical results used to represent $k$-path graphs by restricted normalized sequences in \cite{PEREIRA:14} were presented. The example below illustrates the relation between $k$-path graphs and restricted normalized sequences. 

Consider the 4-path graph $G = P_8 \vee K_3$ of order 11 (see Figure~\ref{4-path_ex}). Its core sequence is given by $\mathcal{F}(G) = \langle F_1, F_2, F_3, F_4, F_5, F_6, F_7\rangle$, with cliques \\
$F_1 = \{ u_1, u_2, u_3, v_1, v_2\}$, $F_2 = \{ u_1, u_2, u_3, v_2, v_3\}$, $F_3 = \{ u_1, u_2, u_3, v_3, v_4\}$, $F_4 = \{ u_1, u_2, u_3, v_4, v_5\}$, 
$F_5 = \{ u_1, u_2, u_3, v_5, v_6\}$, $F_6 = \{ u_1, u_2, u_3, v_6, v_7\}$ e $F_7 =$  $ \{ u_1, u_2, u_3, v_7, v_8\}$.
The sequence $\langle v_3, v_4, v_5, v_6, v_7, v_8 \rangle$ is composed of vertices in the difference of sets representing consecutive 5-cliques in $\mathcal{F}(G)$, $\{ v_i \} = F_{i+1} \backslash F_i$, $1 \leq i \leq n - 6$. A restricted normalized sequence of size $6$ and at most $5$ different integers for  $G = P_8 \vee K_3$ is $C = \langle 1, 2, 1, 2, 1, 2 \rangle$. 

\begin{figure}
\begin{center}
\begin{tikzpicture}[scale=0.75, every node/.style={circle, draw, fill=white, inner sep=1pt, minimum size=20pt}]
\foreach \i in {1,...,8} {
    \node (v\i) at (\i*1.2, 0) {$v_{\i}$};
}
\foreach \i in {1,...,7} {
    \pgfmathtruncatemacro{\j}{\i+1}
    \draw (v\i) -- (v\j);
}
\node (u1) at (3.6, 2.5) {$u_1$};
\node (u2) at (5.4, 2.5) {$u_2$};
\node (u3) at (4.5, 3.8) {$u_3$};
\draw (u1) -- (u2);
\draw (u1) -- (u3);
\draw (u2) -- (u3);
\foreach \i in {1,...,8} {
    \foreach \j in {1,2,3} {
        \draw (v\i) -- (u\j);
    }
}
\end{tikzpicture}
\caption{$P_8 \vee K_3$}
\label{4-path_ex}
\end{center}
\end{figure}

\pagebreak

The algorithm that generate color sequences for all non-isomorphic $k$-path graphs of order $n$ initializes with the lexicographically smallest sequence that satisfies the three conditions of Lemma 2 (restricted, normalized and only using positives integers in $\{1,..., k+1\}$). Subsequently, the sequence is modified to produce new sequences. Each generated sequence must be analyzed to verify whether it represents a valid $k$-path graph,  satisfying the required constraints.
Algorithm \ref{alg:algo1} (\cite{PEREIRA:14} ) is presented below whose proof of correctness is given by Lemmas 2 and 3 of Section 2.3.

\begin{algorithm}[H]
\begin{small}
\caption{\\
Input: $n$, $k$.\\
Output: $\mathcal{T}_{k,n}$ set of color sequences of all non-isomorphic $k$-path graphs of order $n$.\\  
}\label{alg:algo1}
\begin{algorithmic}[1]
  \State \textbf{Step 1:}
  \State Set $\langle p_1, p_2, \dots, p_{n-k-1}\rangle \gets \langle 1,2,1,2,\dots\rangle$
  \State $\mathcal{T}_{k,n} \gets \{\langle p_1, p_2, \dots, p_{n-k-1}\rangle\}$
  \State \textbf{Step 2:} \textbf{While} exists maximum index $i$ such that:
    \[
      a_i = \min\bigl(\{p_{i+1},p_{i+2}\}\setminus\{p_{i-1}\}\bigr),
      \quad b_i = \max(p_1,\dots,p_{i-1}) + 1,
      \quad a_i \le b_i \le k+1 \, \textbf{do}
    \] 
  \State \quad \quad \textbf{Step (a)}   $\langle p_1,\dots,p_{n-k-1}\rangle 
      \gets \langle p_1,\dots,p_{i-1},\,a_i,\,1,2,1,2,\dots\rangle$
    
  \State \quad \quad \textbf{Step (b)} Compute
    $\langle q_1,\dots,q_{n-k-1}\rangle$
    the reverse restricted normalized sequence of 
    $\langle p_1,\dots,p_{n-k-1}\rangle$
  \State \quad \quad \textbf{Step (c)}  
  \textbf{If} $\langle p_1,\dots,p_{n-k-1}\rangle \preceq \langle q_1,\dots,q_{n-k-1}\rangle$ \textbf{then}  
   $\mathcal{T}_{k,n} \gets \mathcal{T}_{k,n} \cup \{\langle p_1,\dots,p_{n-k-1}\rangle\}$
   \State \quad\quad\quad\quad\quad\quad\quad\quad\textbf{end if}
  \State \quad\quad\quad\quad \textbf{end while}
  \State {\bf Return} $\mathcal{T}_{k,n}$
\end{algorithmic}
\end{small}
\end{algorithm}

\subsection{Transforming color sequence in $g6$ format - phase ii}\label{detalhes_etapa_b}

In this phase, the color sequences representing $k$-path graphs, with fixed $k$ and $n$, are transformed into $g6$ format, suitable for storage of the lists of graphs. 
This format, initially presented by Brendan McKay, is widely used to store undirected graphs in a compact form, using only printable ASCII characters\footnote{https://users.cecs.anu.edu.au/$\sim$bdm/data/formats.txt}. The sequence of characters is obtained from the order and the columns of the adjacency matrix of the graph. 

First, the color sequence is used to construct the adjacency matrix of the graph. To this end, given  $n$, $k$ and a color sequence $C$ of lenght $n-k-1$ representing a $k$-path graph, Algorithm 2 produces the corresponding adjacency matrix.

\begin{algorithm}[H]
\begin{small}
\caption{\\
Input: $n$, $k$, $C =  \langle c_1, c_2, \dots , c_{n-k-1}\rangle$ .\\
Output: $A(G)$ for $G$, the corresponding $k$-path of order $n$.
}\label{alg:algo2}
\begin{algorithmic}[1]
  \State $B \gets  [1, 2, ..., k+1,  c_1, c_2, \dots , c_{n-k-1}]_{1 \times n}$
  \State  $A(G)  \gets [a_{i,j}]_{n \times n} = \mathbf{0}_{n \times n}$ 
  \State  $\Delta \gets \{1, 2, ..., k+1\}$
  \State \textbf{for} $i=n, n-1, ... ,1$ \textbf{do}  
   \State \quad $Aux \gets \{ B[i]\}$
   \State \quad  \textbf{for} $j= i-1, ... ,1$ \textbf{do}  
   \State \quad \quad \textbf{if}  $ \Delta \not = Aux$ \textbf{then}
   \State \quad \quad \quad \textbf{if}  $B[j] \not \in Aux$ \textbf{then}
   \State \quad \quad \quad \quad $Aux = Aux \cup \{B[j] \}$, $a_{i,j} =  a_{j,i}=1$
   \State {\bf Return} $A(G)$
\end{algorithmic}
\end{small}
\end{algorithm}

For instance, given $n=11$, $k=4$ and $C = \langle 1,2,1,2,1,2 \rangle$, Algorithm 2 computes the elements $a_{i,j}$, begining  from the last row, and returns the following adjacency matrix corresponding to the $4$-path graph of order 11:
\begin{small}
$$ 
A(G) =\left( \begin{array}{ccccccccccc}
0   &  1   &  1  &  1  &  1  &  1 & 1 &  1 & 1 & 1 & 1\\ 
1   &  0   &  1  &  1  &  1  &  1 & 1 &  1 & 1 & 1 & 1 \\
1   &  1   &  0  &  1  &  1  &  1 & 1 &  1 & 1 & 1 & 1 \\
1   &  1   &  1  &  0  &  1  &  0 & 0 & 0  & 0 & 0 & 0\\
1   &  1   &  1  &  1  &  0  &  1 & 0 & 0  & 0 & 0 & 0\\
1   &  1   &  1  &  0  &  1  &  0 & 1 & 0  & 0 & 0 & 0\\
1   &  1   &  1  &  0  &  0  &  1 & 0 & 1  & 0 & 0 & 0\\
1   &  1   &  1  &  0  &  0  &  0 & 1 & 0  & 1 & 0 & 0\\
1   &  1   &  1  &  0  &  0  &  0 & 0 & 1  & 0 & 1 & 0\\
1   &  1   &  1  &  0  &  0  &  0 & 0 & 0  & 1 & 0 & 1\\
1   &  1   &  1  &  0  &  0  &  0 & 0 & 0  & 0 & 1 & 0\\
\end{array} \right).
$$
\end{small}

Finally, from the adjacency matrix obtained before, the $g6$ format can be obtained. The \textit{networkx.Graph} function  \textit{networkx.to\_graph6\_bytes(G).decode('utf-8')} from the $NetworkX$ library allows to execute this transformation.

\subsection{Storing $k$-path graphs in $g6$ format - phase iii}\label{detalhes_etapa_c}

In this phase, the storage of all non-isomorphic $k$-path graphs with fixed $k$ and $n$ in text files is presented. 
For each value of $k = 2, 3, 4$, a text file containing the list of $k$-path graphs in $g6$ format is generated for each bounded $n$. 
Since, for fixed $k$, the number of $k$-path graphs in $\mathcal{T}_{k,n}$ is an exponential function in $n$, a maximum $CPU$ execution time of 1 hour was established for Algorithm \ref{alg:algo1}. Therefore, the largest values considered for the order of the $k$-path graphs with $k = 2, 3, 4$ were $n=26, 19, 18$,  respectively. 

A total of 42 text files were obtained: 21 for 2-path graphs ($6 \leq n \leq 26$), 12 for 3-path graphs ($8 \leq n \leq 19$), and 9 for 4-path graphs ($10 \leq n \leq 18$). 
The Tables \ref{tab:quant_2-caminho}, \ref{tab:quant_3caminho} and \ref{tab:quant_4caminho} show the number of non-isomorphic $k$-path graphs, $k \in \{2, 3, 4\}$, generated for each fixed value of $n$.

\begin{table}[htb]
\begin{tiny}
\centering
\begin{tabular}{|c|r||c|r|}
\hline
$n$ & \# 2-paths of order $n$ & $n$ & \# 2-paths of order $n$ \\ \hline\hline
6  & 2       & 16 & 1.056     \\ \hline
7  & 3       & 17 & 2.080     \\ \hline
8  & 6       & 18 & 4.160     \\ \hline
9  & 10      & 19 & 8.256     \\ \hline
10 & 20      & 20 & 16.512    \\ \hline
11 & 36      & 21 & 32.896    \\ \hline
12 & 72      & 22 & 65.792    \\ \hline
13 & 136     & 23 & 131.328   \\ \hline
14 & 272     & 24 & 262.656   \\ \hline
15 & 528     & 25 & 524.800   \\ \hline
    &         & 26 & 1.049.600  \\ \hline
\end{tabular}
\caption{Number of non-isomorphic 2-paths of order $n$, $6 \leq n \leq 26$.}
\label{tab:quant_2-caminho}
\end{tiny}
\end{table}

\begin{table}[htb]
\begin{tiny}
\centering
\begin{minipage}[b]{0.48\hsize}\centering
\begin{tabular}{|c|r|}
\hline
$n$ & \# 3-paths of order $n$  \\ \hline\hline
8  & 4       \\ \hline
9  & 10      \\ \hline
10 & 25      \\ \hline
11 & 70      \\ \hline
12 & 196     \\ \hline
13 & 574     \\ \hline
14 & 1.681   \\ \hline
15 & 5.002   \\ \hline
16 & 14.884  \\ \hline
17 & 44.530  \\ \hline
18 & 133.225 \\ \hline
19 & 399.310 \\ \hline
\end{tabular}
\caption{Number of non-isomorphic 3-paths, $8 \leq n \leq 19$.}
\label{tab:quant_3caminho}
\end{minipage} 
\hfill
\begin{minipage}[b]{0.48\hsize}\centering
\begin{tabular}{|c|r|}
\hline
$n$ & \# 4-paths of order $n$  \\ \hline\hline
10 & 11       \\ \hline
11 & 31       \\ \hline
12 & 107      \\ \hline
13 & 379      \\ \hline
14 & 1.451    \\ \hline
15 & 5.611    \\ \hline
16 & 22.187   \\ \hline
17 & 87.979   \\ \hline
18 & 350.891  \\ \hline
\end{tabular}
\caption{Number of non-isomorphic 4-paths, $10 \leq n \leq 18$.}
\label{tab:quant_4caminho}
\end{minipage}
\end{tiny}
\end{table}

The lists of $k$-path graphs generated in this work  are available on House of Graphs Meta-directory  \footnote{https://houseofgraphs.org/meta-directory/k-path-graphs} .

\section{Experiments and Results}\label{sec:resulteexperimentos}

This section presents experiments performed with the $k$-path lists generated in Section~\ref{sec:metodologia}, as well as the analysis of the obtained results. The objective now is to determine the $k$-path graphs with fixed $k$ and $n$ in these lists that maximize (minimize) the algebraic connectivity and also those $k$-path graphs that maximize the $\alpha$-index of the $A_{\alpha}$-matrices for $\alpha \in \{ 0.1, ..., 0.9\}$. The experiments in this section were inspired by the work developed by Yu et al. \cite{yu:15}. This article deals with the largest eigenvalue of the signless Laplacian matrix, for particular maximal outerplanar graphs, the 2-path graphs. The authors prove which graphs in this family, when the order is fixed, achieve the maximum and minimum values of the index of the signless Laplacian matrix. Here we look for extremal values for the algebraic connectivity and for the $\alpha$-index.

First, an exhaustive search was conducted on the lists of $k$-path graphs generated in the previous section in order to determine which graphs maximize and minimize algebraic connectivity and maximize the $\alpha$-index. For this tasks, the \texttt{np.linalg.eigvalsh()} function from the \texttt{NumPy} library was used.

\begin{figure}
\centering 
 \includegraphics[width=1.25\textwidth]{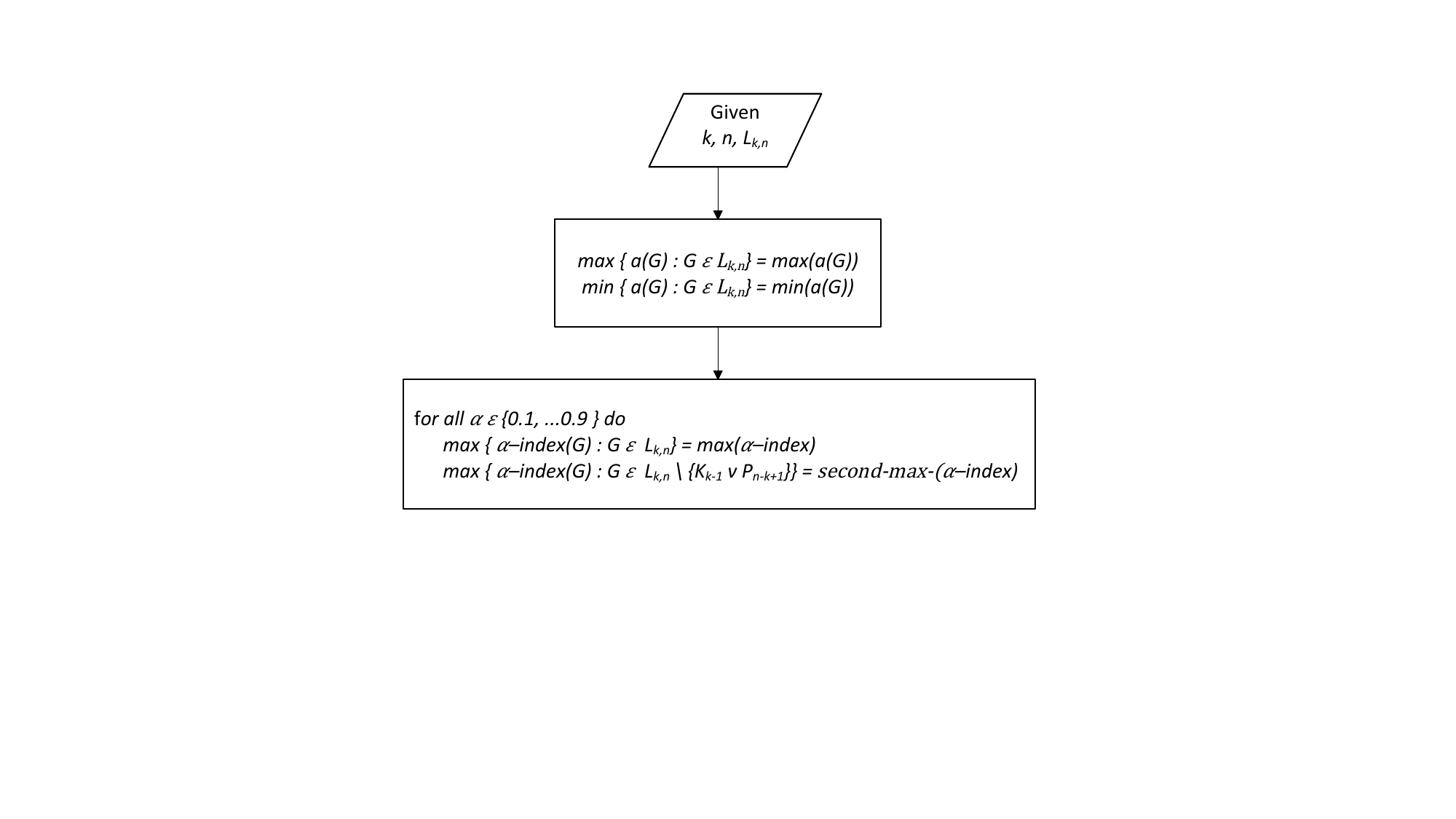}
\vspace{-3.0cm}
 \caption{Exhaustive search}
 \label{exhaustive_search}
\end{figure}

After that, given $k$, the graphs that achieve the maximum and minimum values of the algebraic connectivity in the list $L_{k,n}$ of all $k$-path graphs of fixed order $n$, denoted max($a(G)$) and min($a(G)$), respectively,  were identified.
For the $\alpha$-index, a similar process was performed to obtain the graph that achieves the maximum value in the list $L_{k,n}$, denoted max($\alpha$-index), for each $\alpha \in \{0.1, ..., 0.9\}$. 
Furthermore, the maximum $\alpha$-index value in the list $L_{k,n} \backslash \{K_{k-1} \vee P_{n-k+1}\}$, denoted second-max($\alpha$-index), was also analyzed.
Figure~\ref{exhaustive_search} ilustrates the exhaustive search method.\newline

\subsection{Experiments for algebraic connectivity}\label{expalgebraic}

Tables~\ref{table:2-path-max} and~\ref{table:2-path-min} present the results obtained for the maximum and minimum algebraic connectivity, respectively, of the $2$-path graphs with a fixed number of vertices, ranging from 6 to 26. The $2$-path graphs, in format $g6$ and their respective color sequences, that reach each of the extreme values, were identified. Note that the color sequences  $\langle 1, 2, 1, 2, 1, 2, .... \rangle$ and $\langle 1, 2, 3, 1, 2, 3, 1, 2, 3...\rangle$ correspond to the graphs $K_1 \vee P_{n-1}$ and $P_n^2$, for each fixed $n$, respectively. Therefore, in all experiments performed, the $2$-path graph of fixed order $n$ that maximizes (minimizes) the algebraic connectivity is $K_1 \vee P_{n-1}$ ($P_n^2$).

\begin{table}[htb]
\tiny
\begin{tabular}{|p{0.03\textwidth}|p{0.06\textwidth}|p{0.5\textwidth}|p{0.28\textwidth}|}
\hline
\textbf{$n$} & \textbf{max $a(G)$ } & \textbf{2-path $G$ in $g6$ format} & \textbf{$C$, color sequence for $G$} \\ \hline
6 & 1.3820 & \texttt{EzKg} & 1 2 1 \\ \hline
7 & 1.2679 & \texttt{FzKhG} & 1 2 1 2 \\ \hline
8 & 1.1981 & \texttt{GzKhHC} & 1 2 1 2 1 \\ \hline
9 & 1.1522 & \texttt{HzKhHC`} & 1 2 1 2 1 2 \\ \hline
10 & 1.1206 & \texttt{IzKhHC`GG} & 1 2 1 2 1 2 1 \\ \hline
11 & 1.0979 & \texttt{JzKhHC`GH?\_} & 1 2 1 2 1 2 1 2 \\ \hline
12 & 1.0810 & \texttt{KzKhHC`GH?c@} & 1 2 1 2 1 2 1 2 1 \\ \hline
13 & 1.0681 & \texttt{LzKhHC`GH?c@G@} & 1 2 1 2 1 2 1 2 1 2 \\ \hline
14 & 1.0581 & \texttt{MzKhHC`GH?c@G@G?\_} & 1 2 1 2 1 2 1 2 1 2 1 \\ \hline
15 & 1.0501 & \texttt{NzKhHC`GH?c@G@G?c?G} & 1 2 1 2 1 2 1 2 1 2 1 2 \\ \hline
16 & 1.0437 & \texttt{OzKhHC`GH?c@G@G?c?H?@} & 1 2 1 2 1 2 1 2 1 2 1 2 1 \\ \hline
17 & 1.0384 & \texttt{PzKhHC`GH?c@G@G?c?H?@G?C} & 1 2 1 2 1 2 1 2 1 2 1 2 1 2 \\ \hline
18 & 1.0341 & \texttt{QzKhHC`GH?c@G@G?c?H?@G?C\_?G} & 1 2 1 2 1 2 1 2 1 2 1 2 1 2 1 \\ \hline
19 & 1.0304 & \texttt{RzKhHC`GH?c@G@G?c?H?@G?C\_?H??G} & 1 2 1 2 1 2 1 2 1 2 1 2 1 2 1 2 \\ \hline
20 & 1.0273 & \texttt{SzKhHC`GH?c@G@G?c?H?@G?C\_?H??H??C} & 1 2 1 2 1 2 1 2 1 2 1 2 1 2 1 2 1 \\ \hline
21 & 1.0246 & \texttt{TzKhHC`GH?c@G@G?c?H?@G?C\_?H??H??C\_?@} & 1 2 1 2 1 2 1 2 1 2 1 2 1 2 1 2 1 2 \\ \hline
22 & 1.0223 & \texttt{UzKhHC`GH?c@G@G?c?H?@G?C\_?H??H??C\_?@G??G} & 1 2 1 2 1 2 1 2 1 2 1 2 1 2 1 2 1 2 1 \\ \hline
23 & 1.0204 & \texttt{VzKhHC`GH?c@G@G?c?H?@G?C\_?H??H??C\_?@G??H???\_} & 1 2 1 2 1 2 1 2 1 2 1 2 1 2 1 2 1 2 1 2 \\ \hline
24 & 1.0186 & \texttt{WzKhHC`GH?c@G@G?c?H?@G?C\_?H??H??C\_?@G??H???c??@} & 1 2 1 2 1 2 1 2 1 2 1 2 1 2 1 2 1 2 1 2 1 \\ \hline
25 & 1.0171 & \texttt{XzKhHC`GH?c@G@G?c?H?@G?C\_?H??H??C\_?@G??H???c??@G??@} & 1 2 1 2 1 2 1 2 1 2 1 2 1 2 1 2 1 2 1 2 1 2 \\ \hline
26 & 1.0158 & \texttt{YzKhHC`GH?c@G@G?c?H?@G?C\_?H??H??C\_?@G??H???c??@G??@G???\_} & 1 2 1 2 1 2 1 2 1 2 1 2 1 2 1 2 1 2 1 2 1 2 1 \\ \hline
\end{tabular}
\caption{2-path graph in $g6$ format and its color sequence which maximizes $a(G)$ in $L_{2,n}$.}
\label{table:2-path-max}
\end{table}

\begin{table}[htb]
\tiny
\begin{tabular}{|p{0.03\textwidth}|p{0.06\textwidth}|p{0.5\textwidth}|p{0.28\textwidth}|}
\hline
\textbf{$n$} & \textbf{min $a(G)$} & \textbf{2-path $G$ in $g6$ format} & \textbf{$C$, color sequence for $G$} \\ \hline
6 & 1.1864 & \texttt{EzKW} & 1 2 3 \\ \hline
7 & 0.9139 & \texttt{FzKWW} & 1 2 3 1 \\ \hline
8 & 0.7186 & \texttt{GzKWWK} & 1 2 3 1 2 \\ \hline
9 & 0.5778 & \texttt{HzKWWKB} & 1 2 3 1 2 3 \\ \hline
10 & 0.4735 & \texttt{IzKWWKB?W} & 1 2 3 1 2 3 1 \\ \hline
11 & 0.3946 & \texttt{JzKWWKB?W@\_} & 1 2 3 1 2 3 1 2 \\ \hline
12 & 0.3335 & \texttt{KzKWWKB?W@\_B} & 1 2 3 1 2 3 1 2 3 \\ \hline
13 & 0.2855 & \texttt{LzKWWKB?W@\_B?B} & 1 2 3 1 2 3 1 2 3 1 \\ \hline
14 & 0.2470 & \texttt{MzKWWKB?W@\_B?B?@\_} & 1 2 3 1 2 3 1 2 3 1 2 \\ \hline
15 & 0.2158 & \texttt{NzKWWKB?W@\_B?B?@\_?W} & 1 2 3 1 2 3 1 2 3 1 2 3 \\ \hline
16 & 0.1901 & \texttt{OzKWWKB?W@\_B?B?@\_?W?B} & 1 2 3 1 2 3 1 2 3 1 2 3 1 \\ \hline
17 & 0.1687 & \texttt{PzKWWKB?W@\_B?B?@\_?W?B??K} & 1 2 3 1 2 3 1 2 3 1 2 3 1 2 \\ \hline
18 & 0.1507 & \texttt{QzKWWKB?W@\_B?B?@\_?W?B??K??W} & 1 2 3 1 2 3 1 2 3 1 2 3 1 2 3 \\ \hline
19 & 0.1354 & \texttt{RzKWWKB?W@\_B?B?@\_?W?B??K??W??W} & 1 2 3 1 2 3 1 2 3 1 2 3 1 2 3 1 \\ \hline
20 & 0.1223 & \texttt{SzKWWKB?W@\_B?B?@\_?W?B??K??W??W??K} & 1 2 3 1 2 3 1 2 3 1 2 3 1 2 3 1 2 \\ \hline
21 & 0.1110 & \texttt{TzKWWKB?W@\_B?B?@\_?W?B??K??W??W??K??B} & 1 2 3 1 2 3 1 2 3 1 2 3 1 2 3 1 2 3 \\ \hline
22 & 0.1012 & \texttt{UzKWWKB?W@\_B?B?@\_?W?B??K??W??W??K??B???W} & 1 2 3 1 2 3 1 2 3 1 2 3 1 2 3 1 2 3 1 \\ \hline
23 & 0.0927 & \texttt{VzKWWKB?W@\_B?B?@\_?W?B??K??W??W??K??B???W??@\_} & 1 2 3 1 2 3 1 2 3 1 2 3 1 2 3 1 2 3 1 2 \\ \hline
24 & 0.0852 & \texttt{WzKWWKB?W@\_B?B?@\_?W?B??K??W??W??K??B???W??@\_??B} & 1 2 3 1 2 3 1 2 3 1 2 3 1 2 3 1 2 3 1 2 3 \\ \hline
25 & 0.0785 & \texttt{XzKWWKB?W@\_B?B?@\_?W?B??K??W??W??K??B???W??@\_??B???B} & 1 2 3 1 2 3 1 2 3 1 2 3 1 2 3 1 2 3 1 2 3 1 \\ \hline
26 & 0.0726 & \texttt{YzKhGKD@G@\_D?B?@\_?g?H??K??g??W??S??H???W??A\_??B???B???@\_} & 1 2 3 1 2 3 1 2 3 1 2 3 1 2 3 1 2 3 1 2 3 1 2 \\ \hline
\end{tabular}
\caption{2-path graph in $g6$ format and its color sequence which minimizes $a(G)$ in $L_{2,n}$.}
\label{table:2-path-min}
\end{table}

Furthermore, the corresponding graph lists were used to exhaustively determine which $k$-path graphs for $k=3, 4$ have the maximum and minimum algebraic connectivity for a fixed $n$. In the Appendix, Tables~\ref {table:3-path-max}, \ref{table:3-path-min}, \ref{table:4-path-max} and ~\ref{table:4-path-min} present the maximum and minimum algebraic connectivity value for each fixed $n$, ranging from 8 to 19 (for $k=3$) and from 10 to 18 (for $k=4$). Note that the color sequences $\langle 1, \, 2, \dots 1, \, 2, ...\rangle$ and $\langle 1, \, 2,  \dots k+1,\dots \, 1, \, 2,\dots k+1, ...\rangle$ correspond to the $k$-path graphs $K_{k-1} \vee P_{n-k+1}$ and $P_n^k$ . Therefore, in all experiments performed, the $3$-path and $4$-path graph of fixed order $n$ that maximizes (minimizes) the algebraic connectivity are $K_{k-1} \vee P_{n- k+1}$ ($P_n^k$).

\subsection{Experiments for the eigenvalues of $A_{\alpha}(G)$}

In this section, we present results on the behavior of the 
maximum $\alpha$-index in the lists $L_{k,n}$ and $L_{k,n} \backslash \{K_{k-1} \vee P_{n-k+1}\}$ of $k$-path graphs, $k=2,3,4$, of fixed order $n$.  
The experiments will be executed for values of $\alpha \in \{0.1, 0.2, ..., 0.9\}$.

\subsubsection{The  maximum  $\alpha$-index}\label{expalphaindex}

Tables~\ref{tab:extremal_common} and ~\ref{tab:maximos_Q_Alpha} present the graphs, in $g6$ format and their respective color sequences, that have the maximum $\alpha$-index (for $\alpha \in \{ 0.1, \dots, 0.9\}$) among the $2$-path graphs of fixed order $n$ ($L_{2,n}$). Each row in the tables represents the order of the graph. From Table~\ref{tab:extremal_common} can be concluded that, for all considered values of $\alpha$, the color sequence that corresponds to the graph $K_1 \vee P_{n-1}$ maximizes the $\alpha$-index of the 2-path graphs of fixed order. Table~\ref{tab:maximos_Q_Alpha} shows the maximum values of $\alpha$-index, achieved by $K_1 \vee P_{n-1}$.

As $Q(G)  = 2 A_{1/2}(G)$, Tables ~\ref{tab:extremal_common} and ~\ref{tab:maximos_Q_Alpha} also show $q_1(G) = 2 \lambda_1(A_{1/2}(G))$. Then, the same extremal graphs were identified for the signless Laplacian index for $k$-path graphs of order $n$, with $k=2, 3, 4$.

\begin{table}[ht]
  \centering
  \scriptsize
  \begin{tabular}{|c|c|c|c|c|c|c|c|c|c|l|}
    \hline
    $n$  & $A_{0.1}$ & $A_{0.2}$ & $A_{0.3}$ & $A_{0.4}$ & $A_{0.5}$ & $A_{0.6}$ & $A_{0.7}$ & $A_{0.8}$ & $A_{0.9}$ & Color sequence \\ \hline
    6  & \multicolumn{9}{|l|}{\texttt{\detokenize{EzKg}}} & 1 2 1\\ \hline
    7  & \multicolumn{9}{|l|}{\texttt{\detokenize{FzKhG}}} & 1 2 1 2\\ \hline
    8  & \multicolumn{9}{|l|}{\texttt{\detokenize{GzKhHC}}} & 1 2 1 2 1 \\ \hline
    9  & \multicolumn{9}{|l|}{\texttt{\detokenize{HzKhHC`}}} & 1 2 1 2 1 2 \\ \hline
    10 & \multicolumn{9}{|l|}{\texttt{\detokenize{IzKhHC`GG}}} & 1 2 1 2 1 2 1 \\ \hline
    11 & \multicolumn{9}{|l|}{\texttt{\detokenize{JzKhHC`GH?_}}} & 1 2 1 2 1 2 1 2 \\ \hline
    12 & \multicolumn{9}{|l|}{\texttt{\detokenize{KzKhHC`GH?c@}}} & 1 2 1 2 1 2 1 2 1 \\ \hline
    13 & \multicolumn{9}{|l|}{\texttt{\detokenize{LzKhHC`GH?c@G@}}} & 1 2 1 2 1 2 1 2 1 2 \\ \hline
    14 & \multicolumn{9}{|l|}{\texttt{\detokenize{MzKhHC`GH?c@G@G?_}}} & 1 2 1 2 1 2 1 2 1 2 1 \\ \hline
    15 & \multicolumn{9}{|l|}{\texttt{\detokenize{NzKhHC`GH?c@G@G?c?G}}} & 1 2 1 2 1 2 1 2 1 2 1 2 \\ \hline
    16 & \multicolumn{9}{|l|}{\texttt{\detokenize{OzKhHC`GH?c@G@G?c?H?@}}} & 1 2 1 2 1 2 1 2 1 2 1 2 1 \\ \hline
    17 & \multicolumn{9}{|l|}{\texttt{\detokenize{PzKhHC`GH?c@G@G?c?H?@G?C}}} & 1 2 1 2 1 2 1 2 1 2 1 2 1 2\\ \hline
    18 & \multicolumn{9}{|l|}{\texttt{\detokenize{QzKhHC`GH?c@G@G?c?H?@G?C_?G}}} & 1 2 1 2 1 2 1 2 1 2 1 2 1 2 1 \\ \hline
    19 & \multicolumn{9}{|l|}{\texttt{\detokenize{RzKhHC`GH?c@G@G?c?H?@G?C_?H??G}}} & 1 2 1 2 1 2 1 2 1 2 1 2 1 2 1 2 \\ \hline
    20 & \multicolumn{9}{|l|}{\texttt{\detokenize{SzKhHC`GH?c@G@G?c?H?@G?C_?H??H??C}}} & 1 2 1 2 1 2 1 2 1 2 1 2 1 2 1 2 1\\ \hline
    21 & \multicolumn{9}{|l|}{\texttt{\detokenize{TzKhHC`GH?c@G@G?c?H?@G?C_?H??H??C_?@}}} & 1 2 1 2 1 2 1 2 1 2 1 2 1 2 1 2 1 2\\ \hline
    22 & \multicolumn{9}{|l|}{\texttt{\detokenize{UzKhHC`GH?c@G@G?c?H?@G?C_?H??H??C_?@G??G}}} & 1 2 1 2 1 2 1 2 1 2 1 2 1 2 1 2 1 2 1\\ \hline
    23 & \multicolumn{9}{|l|}{\texttt{\detokenize{VzKhHC`GH?c@G@G?c?H?@G?C_?H??H??C_?@G??H???_}}} & 1 2 1 2 1 2 1 2 1 2 1 2 1 2 1 2 1 2 1 2\\ \hline
    24 & \multicolumn{9}{|l|}{\texttt{\detokenize{WzKhHC`GH?c@G@G?c?H?@G?C_?H??H??C_?@G??H???c??@}}} & 1 2 1 2 1 2 1 2 1 2 1 2 1 2 1 2 1 2 1 2 1\\ \hline
    25 & \multicolumn{9}{|l|}{\texttt{\detokenize{XzKhHC`GH?c@G@G?c?H?@G?C_?H??H??C_?@G??H???c??@G??@}}} & 1 2 1 2 1 2 1 2 1 2 1 2 1 2 1 2 1 2 1 2 1 2\\ \hline
    26 & \multicolumn{9}{|l|}{\texttt{\detokenize{YzKhHC`GH?c@G@G?c?H?@G?C_?H??H??C_?@G??H???c??@G??@G???_}}} & 1 2 1 2 1 2 1 2 1 2 1 2 1 2 1 2 1 2 1 2 1 2 1 \\ \hline
  \end{tabular}
  \caption{$2$-paths with maximum $\alpha$-index in $L_{2,n}$. The structure of the extremal graph is the same for all $\alpha \in \{0.1,\dots,0.9\}$.}
  \label{tab:extremal_common}
\end{table}

\begin{table}[H]
\scriptsize
\centering
\begin{tabular}{|c|c|c|c|c|c|c|c|c|c|}
\hline
 & max & max & max & max & max & max & max & max &  \\
$n$ & ${0.1}$-index & ${0.2}$-index & ${0.3}$-index & ${0.4}$-index & ${0.5}$-index & ${0.6}$-index & ${0.7}$-index & ${0.8}$-index & ${0.9}$-index \\
\hline
6  & 3.2494 & 3.2835 & 3.3284 & 3.3900 & 3.4788 & 3.6127 & 3.8192 & 4.1237 & 4.5260 \\\hline
7  & 3.5068 & 3.5634 & 3.6401 & 3.7479 & 3.9050 & 4.1365 & 4.4670 & 4.9016 & 5.4215 \\\hline
8  & 3.7394 & 3.8242 & 3.9410 & 4.1070 & 4.3462 & 4.6850 & 5.1364 & 5.6897 & 6.3191 \\\hline
9  & 3.9555 & 4.0733 & 4.2373 & 4.4703 & 4.8003 & 5.2490 & 5.8167 & 6.4822 & 7.2177 \\\hline
10 & 4.1597 & 4.3145 & 4.5314 & 4.8384 & 5.2641 & 5.8227 & 6.5030 & 7.2771 & 8.1166 \\\hline
11 & 4.3547 & 4.5499 & 4.8247 & 5.2108 & 5.7352 & 6.4029 & 7.1929 & 8.0734 & 9.0159 \\\hline
12 & 4.5422 & 4.7810 & 5.1176 & 5.5868 & 6.2117 & 6.9874 & 7.8853 & 8.8706 & 9.9153 \\\hline
13 & 4.7237 & 5.0086 & 5.4106 & 5.9660 & 6.6922 & 7.5750 & 8.5792 & 9.6684 & 10.8149 \\\hline
14 & 4.8999 & 5.2334 & 5.7038 & 6.3477 & 7.1759 & 8.1649 & 9.2743 & 10.4666 & 11.7145 \\\hline
15 & 5.0717 & 5.4558 & 5.9973 & 6.7317 & 7.6621 & 8.7565 & 9.9703 & 11.2652 & 12.6142 \\\hline
16 & 5.2395 & 5.6763 & 6.2911 & 7.1175 & 8.1502 & 9.3494 & 10.6669 & 12.0639 & 13.5140 \\\hline
17 & 5.4038 & 5.8950 & 6.5852 & 7.5049 & 8.6400 & 9.9434 & 11.3640 & 12.8629 & 14.4138 \\\hline
18 & 5.5650 & 6.1123 & 6.8797 & 7.8936 & 9.1310 & 10.5381 & 12.0616 & 13.6620 & 15.3136 \\\hline
19 & 5.7234 & 6.3283 & 7.1745 & 8.2834 & 9.6230 & 11.1335 & 12.7594 & 14.4612 & 16.2134 \\\hline
20 & 5.8793 & 6.5432 & 7.4696 & 8.6742 & 10.1160 & 11.7295 & 13.4575 & 15.2605 & 17.1133 \\\hline
21 & 6.0328 & 6.7571 & 7.7650 & 9.0659 & 10.6097 & 12.3259 & 14.1559 & 16.0599 & 18.0132 \\\hline
22 & 6.1843 & 6.9700 & 8.0607 & 9.4583 & 11.1041 & 12.9227 & 14.8544 & 16.8594 & 18.9130 \\\hline
23 & 6.3337 & 7.1822 & 8.3566 & 9.8514 & 11.5990 & 13.5199 & 15.5530 & 17.6589 & 19.8129 \\\hline
24 & 6.4814 & 7.3937 & 8.6528 & 10.2450 & 12.0944 & 14.1173 & 16.2518 & 18.4584 & 20.7129 \\\hline
25 & 6.6274 & 7.6044 & 8.9492 & 10.6391 & 12.5902 & 14.7149 & 16.9507 & 19.2581 & 21.6128 \\\hline
26 & 6.7718 & 7.8146 & 9.2457 & 11.0337 & 13.0863 & 15.3128 & 17.6497 & 20.0577 & 22.5127 \\
\hline
\end{tabular}
\caption{Maximum $\alpha$-index ($\alpha \in \{0.1,\dots,0.9\}$) for 2-paths in $L_{2,n}$,  $6 \leq n \leq 26$.}
\label{tab:maximos_Q_Alpha}
\end{table}

The same experiments were performed for 3-path and 4-path graphs. The Appendix shows Tables~\ref{tab:extremal_common3} and~\ref{tab:maximos_Q_Alpha3} with the results for 3-path graphs and Tables~\ref{tab:extremal_common4} and~\ref{tab:maximos_Q_Alpha4} for 4-path graphs. In all experiments, for all considered values of $n$ and $\alpha$, the color sequence related to the graph that achieves the maximum $\alpha$-index is the sequence $\langle 1 \, 2 \dots 1 \, 2 ...\rangle$, which corresponds to the graph $K_2 \vee P_{n-2}$ for 3-paths in $L_{3,n}$ and $K_3 \vee P_{n-3}$ for 4-paths in $L_{4,n}$ .

\subsubsection{The second maximum $\alpha$-index}\label{expsecondalphaindex}

In addition to analyzing the maximum $\alpha$-index of the $k$-path graphs with $k = 2,3,4$ in $L_{k,n}$,  
it was considered to evaluate the maximum $\alpha$-index in $L_{k,n} \backslash \{K_{k-1} \vee P_{n-k+1}\}$ 
for $\alpha \in \{ 0.1, \dots, 0.9\}$.

Tables~\ref{tab:extremal_common2alpha} and ~\ref{tab:maximos_Q_Alpha2alpha} present the graphs, in $g6$ format and their respective color sequences, that have the maximum second-max($\alpha$-index) 
for $\alpha \in \{ 0.1, \dots, 0.9\}$ among the $2$-path graphs of fixed order $n$ in $L_{2,n}$. Each row in the tables represents the order of the graph. From Table~\ref{tab:extremal_common2alpha} can be concluded that, for all considered values of $\alpha$, the color sequences of size $n-k+1$ given by $\langle 1,\,  2,\, \dots, 1,\, 2,\, \dots, 1,\, 2,\, 3\rangle$ or $\langle 1,\, 2,\, 1,\, 2, .... 1,\, 3 \rangle$,
according to the parity of $n$, 
 correspond to the graph that have maximum second-max($\alpha$-index). Table~\ref{tab:maximos_Q_Alpha2alpha} shows the maximum values of 
second-max($\alpha$-index) among all 2-path graphs of fixed order $n$ in $L_{2,n}$ forall $\alpha \in \{ 0.1, \dots, 0.9\}$.  

\begin{table}[H]
  \centering
  \scriptsize
  \begin{tabular}{|c|c|c|c|c|c|c|c|c|c|l|}
    \hline
    $n$ &  $A_{0.1}$ & $A_{0.2}$ & $A_{0.3}$ & $A_{0.4}$ & $A_{0.5}$ & $A_{0.6}$ & $A_{0.7}$ & $A_{0.8}$ & $A_{0.9}$ & Color sequence  \\ \hline
    6 & \multicolumn{9}{|l|}{\texttt{\detokenize{EzKW}}} & 1 2 3 \\ \hline
    7 & \multicolumn{9}{|l|}{\texttt{\detokenize{FzKgW}}} & 1 2 1 3 \\ \hline
    8 & \multicolumn{9}{|l|}{\texttt{\detokenize{GzKhGK}}} & 1 2 1 2 3 \\ \hline
    9 & \multicolumn{9}{|l|}{\texttt{\detokenize{HzKhHCB}}} & 1 2 1 2 1 3 \\ \hline
    10 & \multicolumn{9}{|l|}{\texttt{\detokenize{IzKhHC`?W}}} & 1 2 1 2 1 2 3 \\ \hline
    11 & \multicolumn{9}{|l|}{\texttt{\detokenize{JzKhHC`GG@_}}} & 1 2 1 2 1 2 1 3 \\ \hline
    12 & \multicolumn{9}{|l|}{\texttt{\detokenize{KzKhHC`GH?_B}}} & 1 2 1 2 1 2 1 2 3 \\ \hline
    13 & \multicolumn{9}{|l|}{\texttt{\detokenize{LzKhHC`GH?c@?B}}} & 1 2 1 2 1 2 1 2 1 3 \\ \hline
    14 & \multicolumn{9}{|l|}{\texttt{\detokenize{MzKhHC`GH?c@G@?@_}}} & 1 2 1 2 1 2 1 2 1 2 3 \\ \hline
    15 & \multicolumn{9}{|l|}{\texttt{\detokenize{NzKhHC`GH?c@G@G?_?W}}} & 1 2 1 2 1 2 1 2 1 2 1 3 \\ \hline
    16 & \multicolumn{9}{|l|}{\texttt{\detokenize{OzKhHC`GH?c@G@G?c?G?B}}} & 1 2 1 2 1 2 1 2 1 2 1 2 3 \\ \hline
    17 & \multicolumn{9}{|l|}{\texttt{\detokenize{PzKhHC`GH?c@G@G?c?H?@??K}}} & 1 2 1 2 1 2 1 2 1 2 1 2 1 3 \\ \hline
    18 & \multicolumn{9}{|l|}{\texttt{\detokenize{QzKhHC`GH?c@G@G?c?H?@G?C??W}}} & 1 2 1 2 1 2 1 2 1 2 1 2 1 2 3 \\ \hline
    19 & \multicolumn{9}{|l|}{\texttt{\detokenize{RzKhHC`GH?c@G@G?c?H?@G?C_?G??W}}} & 1 2 1 2 1 2 1 2 1 2 1 2 1 2 1 3 \\ \hline
    20 & \multicolumn{9}{|l|}{\texttt{\detokenize{SzKhHC`GH?c@G@G?c?H?@G?C_?H??G??K}}} & 1 2 1 2 1 2 1 2 1 2 1 2 1 2 1 2 3 \\ \hline
    21 & \multicolumn{9}{|l|}{\texttt{\detokenize{TzKhHC`GH?c@G@G?c?H?@G?C_?H??H??C??B}}} & 1 2 1 2 1 2 1 2 1 2 1 2 1 2 1 2 1 3 \\ \hline
    22 & \multicolumn{9}{|l|}{\texttt{\detokenize{UzKhHC`GH?c@G@G?c?H?@G?C_?H??H??C_?@???W}}} & 1 2 1 2 1 2 1 2 1 2 1 2 1 2 1 2 1 2 3 \\ \hline
    23 & \multicolumn{9}{|l|}{\texttt{\detokenize{VzKhHC`GH?c@G@G?c?H?@G?C_?H??H??C_?@G??G??@_}}} & 1 2 1 2 1 2 1 2 1 2 1 2 1 2 1 2 1 2 1 3 \\ \hline
    24 & \multicolumn{9}{|l|}{\texttt{\detokenize{WzKhHC`GH?c@G@G?c?H?@G?C_?H??H??C_?@G??H???_??B}}} & 1 2 1 2 1 2 1 2 1 2 1 2 1 2 1 2 1 2 1 2 3 \\ \hline
    25 & \multicolumn{9}{|l|}{\texttt{\detokenize{XzKhHC`GH?c@G@G?c?H?@G?C_?H??H??C_?@G??H???c??@???B}}} & 1 2 1 2 1 2 1 2 1 2 1 2 1 2 1 2 1 2 1 2 1 3 \\ \hline
    26 & \multicolumn{9}{|l|}{\texttt{\detokenize{YzKhHC`GH?c@G@G?c?H?@G?C_?H??H??C_?@G??H???c??@G??@???@_}}} & 1 2 1 2 1 2 1 2 1 2 1 2 1 2 1 2 1 2 1 2 1 2 3 \\ \hline
  \end{tabular}
  \caption{$2$-paths in $L_{2,n}$ with second maximum $\alpha$-index. The structure of the extremal graph is the same for all $\alpha \in \{0.1,\dots,0.9\}$.}
  \label{tab:extremal_common2alpha}
\end{table}

\begin{table}[H]
\scriptsize
\centering
\begin{tabular}{|c|c|c|c|c|c|c|c|c|c|}
\hline
$n$ &  {second} &  {second-} &  {second-} &  {second-} &  {second-} &  {second-} &  {second-} &  {second-} &  {second-} \\
 &  {max} &  {max} &  {max} &  {max} &  {max} &  {max} &  {max} &  {max} &  {max} \\
 &  {${0.1}$-index} &  {$0.2$-index} &  {$0.3$-index} &  {${0.4}$-index} &  {${0.5}$-index} &  {${0.6}$-index} &  {${0.7}$-index} &  {${0.8}$-index} &  {${0.9}$-index} \\

\hline
6  & 3.1996 & 3.2209 & 3.2472 & 3.2802 & 3.3229 & 3.3798 & 3.4588 & 3.5734 & 3.7456 \\ \hline
7  & 3.4208 & 3.4541 & 3.4964 & 3.5521 & 3.6286 & 3.7388 & 3.9060 & 4.1639 & 4.5346 \\ \hline
8  & 3.6317 & 3.6854 & 3.7567 & 3.8548 & 3.9954 & 4.2025 & 4.5046 & 4.9166 & 5.4246 \\ \hline
9  & 3.8352 & 3.9162 & 4.0264 & 4.1815 & 4.4047 & 4.7237 & 5.1567 & 5.6974 & 6.3207 \\ \hline
10 & 4.0319 & 4.1454 & 4.3024 & 4.5247 & 4.8404 & 5.2739 & 5.8292 & 6.4869 & 7.2186 \\ \hline
11 & 4.2225 & 4.3728 & 4.5827 & 4.8796 & 5.2931 & 5.8399 & 6.5114 & 7.2803 & 8.1173 \\ \hline
12 & 4.4074 & 4.5982 & 4.8660 & 5.2429 & 5.7569 & 6.4154 & 7.1990 & 8.0757 & 9.0164 \\ \hline
13 & 4.5874 & 4.8217 & 5.1517 & 5.6124 & 6.2285 & 6.9969 & 7.8898 & 8.8723 & 9.9157 \\ \hline
14 & 4.7629 & 5.0434 & 5.4391 & 5.9868 & 6.7056 & 7.5825 & 8.5828 & 9.6697 & 10.8152 \\ \hline
15 & 4.9343 & 5.2635 & 5.7280 & 6.3650 & 7.1868 & 8.1709 & 9.2772 & 10.4677 & 11.7147 \\ \hline
16 & 5.1022 & 5.4822 & 6.0181 & 6.7462 & 7.6711 & 8.7614 & 9.9726 & 11.2660 & 12.6144 \\ \hline
17 & 5.2667 & 5.6996 & 6.3091 & 7.1299 & 8.1578 & 9.3535 & 10.6689 & 12.0647 & 13.5141 \\ \hline
18 & 5.4284 & 5.9157 & 6.6010 & 7.5155 & 8.6464 & 9.9468 & 11.3657 & 12.8635 & 14.4139 \\ \hline
19 & 5.5873 & 6.1309 & 6.8936 & 7.9028 & 9.1365 & 10.5411 & 12.0630 & 13.6625 & 15.3137 \\ \hline
20 & 5.7437 & 6.3450 & 7.1869 & 8.2915 & 9.6279 & 11.1361 & 12.7606 & 14.4617 & 16.2135 \\ \hline
21 & 5.8979 & 6.5584 & 7.4807 & 8.6814 & 10.1202 & 11.7318 & 13.4586 & 15.2609 & 17.1134 \\ \hline
22 & 6.0500 & 6.7709 & 7.7750 & 9.0723 & 10.6135 & 12.3279 & 14.1568 & 16.0603 & 18.0132 \\ \hline
23 & 6.2001 & 6.9827 & 8.0697 & 9.4640 & 11.1074 & 12.9245 & 14.8552 & 16.8597 & 18.9131 \\ \hline
24 & 6.3484 & 7.1938 & 8.3648 & 9.8565 & 11.6020 & 13.5215 & 15.5538 & 17.6592 & 19.8130 \\ \hline
25 & 6.4951 & 7.4043 & 8.6602 & 10.2496 & 12.0970 & 14.1187 & 16.2525 & 18.4587 & 20.7129 \\ \hline
26 & 6.6401 & 7.6143 & 8.9559 & 10.6433 & 12.5926 & 14.7162 & 16.9514 & 19.2583 & 21.6128 \\ \hline

\end{tabular}
\caption{Maximum ${\alpha}$-index ($\alpha \in \{0.1,\dots,0.9\}$) for 2-paths in $L_{2,n} \backslash K_1 \vee P_{n-1}$, $6 \leq n \leq 26$.}
\label{tab:maximos_Q_Alpha2alpha}
\end{table}

The same experiments were performed for 3-path and 4-path graphs. The Appendix shows Tables~\ref{tab:extremal_common3alpha_2} and ~\ref{tab:maximos_Q_Alpha3alpha_2} with the results for 3-paths and Tables ~\ref{tab:extremal_common4alpha} and ~\ref{tab:maximos_Q_Alpha4alpha} for 4-paths. In all experiments, for all considered values of $n$ and $\alpha$, the color sequence related to the graph that achieves the second-max($\alpha$-index) for 3-path and 4-path graphs has color sequence  $\langle 1, \, 2, \dots 1, \, 2, \, \dots, 1,\, 2,\, 3 \rangle$ or $\langle 1,\, 2,\, 1,\, 2, .... 1,\, 3 \rangle$,
according to the parity of $n$. Tables~\ref{tab:maximos_Q_Alpha3alpha_2} and ~\ref{tab:maximos_Q_Alpha4alpha} show the maximum values of 
second-max($\alpha$-index) among all 3-path and 4-paths of fixed order $n$ in $L_{3,n}$ and $L_{4,n}$, respectively, for all $\alpha$ considered.

\subsection{Conjectures}

In this section, conjectures  based in the results obtained in Subsection~\ref{expalgebraic}, ~\ref{expalphaindex} and ~\ref{expsecondalphaindex} and the structural patterns of $k$-path graphs are formulated.

Since the color sequences for the $k$-path graphs $K_{k-1} \vee P_{n-k+1}$ and $P_{n}^{k}$ are $\langle 1, \, 2, \dots 1, \, 2, ...\rangle$ and $\langle 1, \, 2,  \dots k+1,\dots \, 1, \, 2,\dots k+1, ...\rangle$, respectively, we can present the following conjecture about the  $k$-path graphs with $k \geq 2$ and fixed $n$ that maximize and minimize the algebraic connectivity.

\begin{conj}
Given fixed $n \geq k+1$ and $k \geq 2$, the unique $k$-path graph that maximizes the algebraic connectivity is $K_{k-1} \vee P_{n-k+1}$. Moreover, in the same conditions, the unique  $k$-path graph that minimizes the algebraic connectivity is $P_{n}^{k}$.
\end{conj}

On the other hand, regarding to $A_{\alpha}$-matrix, in all the experiments conducted, the unique $k$-path graph that  maximizes the $\alpha$-index, for all $\alpha \in \{0.1,\dots,0.9\}$ is $K_{k-1} \vee P_{n-k+1}$.

In the case of the second maximum $\alpha$-index, the color sequence obtained in all  the experiments for $k$-path graphs of order $n$, according to its parity, is one of the sequences of size $n-k+1$ given by $\langle 1,\, 2,\, 1,\, 2, .... 1,\, 2,\, 3 \rangle$, $\langle 1,\, 2,\, 1,\, 2, .... 1,\, 3 \rangle$.

The unique $k$-path graph of order $n$ that correspond to this color sequences, according to the parity of $n$, is $G = K_{k-1} \vee P_{n-k+1} - \{e'\} + \{e^*\}$, where $e'$ is and edge between one degree $k$ vertex  adjacent to the universal vertex in $K_{k-1} \vee P_{n-1-k+1}$, denoted $v$, and $e^*$ is an edge between vertex $v$ and the unique vertex $u$ adjacent to $v$ in $P_{n-k+1}$. Here we call this graph \textbf{$k$-weak-generalized-fan}.   
Figures~\ref{2-gen-fan} and~\ref{2-weak-gen-fan} show $K_1 \vee P_6$ and 2-weak-generalized-fan graph of order 7 $K_1 \vee P_6 - (u_7, u_1) + (u_7, u_5)$ , respectively. 

\begin{minipage}{\linewidth}
      \centering
      \begin{minipage}{0.45\linewidth}
      \begin{figure}[H]
\begin{tikzpicture}[scale=0.8, every node/.style={circle, draw, fill=white, inner sep=1pt, minimum size=18pt}]
\node (u1) at (3.6, 2.5) {$u_1$};
\node (u2) at (5.4, 2.5) {$u_4$};
\node (u3) at (4.5, 3.8) {$u_3$};
\node (u0) at (2.7, 3.8) {$u_2$};
\node (u4) at (2.7, 1.2) {$u_6$};
\node (u5) at (4.5, 1.2) {$u_5$};
\node (u6) at (1.8, 2.5) {$u_7$};
\draw (u1) -- (u2);
\draw (u1) -- (u3);
\draw (u2) -- (u3);
\draw (u0) -- (u3);
\draw (u0) -- (u1);
\draw (u4) -- (u1);
\draw (u5) -- (u1);
\draw (u5) -- (u2);
\draw (u4) -- (u5);
\draw (u1) -- (u6);
\draw (u4) -- (u6);
\end{tikzpicture}
\caption{$K_1 \vee P_6$}
\label{2-gen-fan}
      \end{figure}
      \end{minipage}
      \hspace{0.05\linewidth}
      \begin{minipage}{0.45\linewidth}
          \begin{figure}[H]
\begin{tikzpicture}[scale=0.8, every node/.style={circle, draw, fill=white, inner sep=1pt, minimum size=18pt}]
\node (u1) at (3.6, 2.5) {$u_1$};
\node (u2) at (5.4, 2.5) {$u_4$};
\node (u3) at (4.5, 3.8) {$u_3$};
\node (u0) at (2.7, 3.8) {$u_2$};
\node (u4) at (2.7, 1.2) {$u_6$};
\node (u5) at (4.5, 1.2) {$u_5$};
\node (u6) at (3.6, -0.1) {$u_7$};
\draw (u1) -- (u2);
\draw (u1) -- (u3);
\draw (u2) -- (u3);
\draw (u0) -- (u3);
\draw (u0) -- (u1);
\draw (u4) -- (u1);
\draw (u5) -- (u1);
\draw (u5) -- (u2);
\draw (u4) -- (u5);
\draw (u4) -- (u6);
\draw (u5) -- (u6);
\end{tikzpicture}
\caption{$K_1 \vee P_6 - (u_7, u_1) + (u_7, u_5)$}
\label{2-weak-gen-fan}
      \end{figure}
      \end{minipage}
  \end{minipage}

Finally, a conjecture about the maximum $\alpha$-index and the second maximum $\alpha$-index for $k$-path graphs of fixed order $n$ and for all $\alpha \in (0,1)$ is presented.

\begin{conj}
Given fixed $n \geq k+1$ and $k \geq 2$ the unique $k$-path graph that maximizes $\alpha$-index for $\alpha \in (0,1)$ is  $K_{k-1} \vee P_{n-k+1}$. 
Moreover, in the same conditions, the unique $k$-path graph that  attains the second maximum ${\alpha}$-index is the $k$-weak-generalized-fan. 
\end{conj}


\section{Conclusions}\label{sec:concideracoesFinais}

In this work, we present lists of $k$-path graphs for $k \in \{2, 3, 4\}$ and bounded order in $g6$ format. The following ranges for the number of vertices were considered: $6 \leq n \leq 26$ for 2-paths, $8 \leq n \leq 19$ for 3-paths, and $10 \leq n \leq 18$ for 4-paths. These lists were used extensively to obtain information about the behavior of the eigenvalues of the Laplacian and the $A_{\alpha}$-matrices. As a result of this work, given fixed $n \geq k+1$ and a bounded $n$, it was possible to identify the structure of the $k$-path graph that maximizes or minimizes the algebraic connectivity, and maximizes the $\alpha$-index, and the second maximum ${\alpha}$-index, for $\alpha \in \{0.1, ..., 0.9 \}$. Finally, two conjectures were proposed regarding the mentioned eigenvalues.

It is worth mentioning that all the experimental results obtained in this work for 2-path and 3-path graphs are consistent with the theorems and conjectures set in \cite{yu:15} and \cite{Yu:19} for outerplanar and planar graphs. However, 4-path graphs are non-planar, and to our knowledge, do no other works exist in the literature about extremal graphs for the algebraic connectivity or $\alpha$-index of $k$-path graphs with $k \geq 4$.

The obtained results allow identifying research problems to be addressed. For instance, to find a formal proof of the conjectures presented here, as well as the use of the lists of $k$-path graphs to determine new conjectures in Graph Theory and Spectral Graph Theory.

\section{Acknowledgment}
The first author was supported by CEFET/RJ Doctoral Scholarship Program. The second and third authors were partially supported by CNPq, Grant 405552/2023-8 and FAPERJ, Process SEI 260003/001228/2023. The fourth author was supported PIBIC-CNPq Institutional Scientific Iniciation Scholarship, Military Engineering Institute, Brazil.


\section{Appendix}

\subsection{Maximum and minimum  algebraic connectivity of 3-paths and 4-paths of fixed order $n$}

\begin{table}[htb]
\scriptsize
\begin{tabular}{|p{0.03\textwidth}|p{0.06\textwidth}|p{0.5\textwidth}|p{0.28\textwidth}|}
\hline
\textbf{$n$} & \textbf{max $a(G)$} & \textbf{3-path $G$ in $g6$ format} & \textbf{$C$, color sequence for $G$} \\ \hline
8  & 2.2679 & \texttt{G\textasciitilde[xhc} & 1 2 1 2 \\ \hline
9  & 2.1981 & \texttt{H\textasciitilde[xhcp} & 1 2 1 2 1 \\ \hline
10 & 2.1522 & \texttt{I\textasciitilde[xhcpKG} & 1 2 1 2 1 2 \\ \hline
11 & 2.1206 & \texttt{J\textasciitilde[xhcpKH\_\_} & 1 2 1 2 1 2 1 \\ \hline
12 & 2.0979 & \texttt{K\textasciitilde[xhcpKH\_e@} & 1 2 1 2 1 2 1 2 \\ \hline
13 & 2.0810 & \texttt{L\textasciitilde[xhcpKH\_e@K@} & 1 2 1 2 1 2 1 2 1 \\ \hline
14 & 2.0681 & \texttt{M\textasciitilde[xhcpKH\_e@K@K?\_} & 1 2 1 2 1 2 1 2 1 2 \\ \hline
15 & 2.0581 & \texttt{N\textasciitilde[xhcpKH\_e@K@K?e?G} & 1 2 1 2 1 2 1 2 1 2 1 \\ \hline
16 & 2.0501 & \texttt{O\textasciitilde[xhcpKH\_e@K@K?e?H\_@} & 1 2 1 2 1 2 1 2 1 2 1 2 \\ \hline
17 & 2.0437 & \texttt{P\textasciitilde[xhcpKH\_e@K@K?e?H\_@K?C} & 1 2 1 2 1 2 1 2 1 2 1 2 1 \\ \hline
18 & 2.0384 & \texttt{Q\textasciitilde[xhcpKH\_e@K@K?e?H\_@K?Co?G} & 1 2 1 2 1 2 1 2 1 2 1 2 1 2 \\ \hline
19 & 2.0341 & \texttt{R\textasciitilde[xhcpKH\_e@K@K?e?H\_@K?Co?H\_?G} & 1 2 1 2 1 2 1 2 1 2 1 2 1 2 1 \\ \hline
\end{tabular}

\caption{3-path graph, in $g6$ format and its color sequence, which maximizes $a(G)$ in $L_{3,n}$.}
\label{table:3-path-max}
\end{table}

\begin{table}[htb]
\scriptsize
\begin{tabular}{|p{0.03\textwidth}|p{0.06\textwidth}|p{0.5\textwidth}|p{0.28\textwidth}|}
\hline
\textbf{$n$} & \textbf{min $a(G)$} & \textbf{3-path $G$ in $g6$ format} & \textbf{$C$, color sequence for $G$ } \\ \hline
8  & 1.7926 & \texttt{G\textasciitilde[ww[} & 1 2 3 4 \\ \hline
9  & 1.4854 & \texttt{H\textasciitilde[ww[F} & 1 2 3 4 1 \\ \hline
10 & 1.2430 & \texttt{I\textasciitilde[ww[F?w} & 1 2 3 4 1 2 \\ \hline
11 & 1.0502 & \texttt{J\textasciitilde[ww[F?wB\_} & 1 2 3 4 1 2 3 \\ \hline
12 & 0.8968 & \texttt{K\textasciitilde[ww[F?wB\_F} & 1 2 3 4 1 2 3 4 \\ \hline
13 & 0.7733 & \texttt{L\textasciitilde[ww[F?wB\_F?F} & 1 2 3 4 1 2 3 4 1 \\ \hline
14 & 0.6729 & \texttt{M\textasciitilde[ww[F?wB\_F?F?B\_} & 1 2 3 4 1 2 3 4 1 2 \\ \hline
15 & 0.5903 & \texttt{N\textasciitilde[ww[F?wB\_F?F?B\_?w} & 1 2 3 4 1 2 3 4 1 2 3 \\ \hline
16 & 0.5217 & \texttt{O\textasciitilde[ww[F?wB\_F?F?B\_?w?F} & 1 2 3 4 1 2 3 4 1 2 3 4 \\ \hline
17 & 0.4642 & \texttt{P\textasciitilde[ww[F?wB\_F?F?B\_?w?F??[} & 1 2 3 4 1 2 3 4 1 2 3 4 1 \\ \hline
18 & 0.4156 & \texttt{Q\textasciitilde[ww[F?wB\_F?F?B\_?w?F??[??w} & 1 2 3 4 1 2 3 4 1 2 3 4 1 2 \\ \hline
19 & 0.3742 & \texttt{R\textasciitilde[ww[F?wB\_F?F?B\_?w?F??[??w??w} & 1 2 3 4 1 2 3 4 1 2 3 4 1 2 3 \\ \hline
\end{tabular}

\caption{3-path graph, in $g6$ format and its color sequence, which minimizes $a(G)$ in $L_{3,n}$.}
\label{table:3-path-min}
\end{table}

\begin{table}[htb]
\scriptsize
\begin{tabular}{|p{0.03\textwidth}|p{0.06\textwidth}|p{0.5\textwidth}|p{0.28\textwidth}|}
\hline
\textbf{$n$} & \textbf{max $a(G)$} & \textbf{4-path $G$ in $g6$ format} & \textbf{$C$, color sequence for $G$} \\ \hline
10 & 3.1981 & \texttt{I\textasciitilde|xxsxMG} & 1 2 1 2 1 \\ \hline
11 & 3.1522 & \texttt{J\textasciitilde|xxsxMHo\_} & 1 2 1 2 1 2 \\ \hline
12 & 3.1206 & \texttt{K\textasciitilde|xxsxMHof@} & 1 2 1 2 1 2 1 \\ \hline
13 & 3.0979 & \texttt{L\textasciitilde|xxsxMHof@M@} & 1 2 1 2 1 2 1 2 \\ \hline
14 & 3.0810 & \texttt{M\textasciitilde|xxsxMHof@M@M?{\_}} & 1 2 1 2 1 2 1 2 1 \\ \hline
15 & 3.0681 & \texttt{N\textasciitilde|xxsxMHof@M@M?f?G} & 1 2 1 2 1 2 1 2 1 2 \\ \hline
16 & 3.0581 & \texttt{O\textasciitilde|xxsxMHof@M@M?f?Ho@} & 1 2 1 2 1 2 1 2 1 2 1 \\ \hline
17 & 3.0501 & \texttt{P\textasciitilde|xxsxMHof@M@M?f?Ho@M?C} & 1 2 1 2 1 2 1 2 1 2 1 2 \\ \hline
18 & 3.0437 & \texttt{Q\textasciitilde|xxsxMHof@M@M?f?Ho@M?Cw?G} & 1 2 1 2 1 2 1 2 1 2 1 2 1 \\ \hline
\end{tabular}

\caption{4-path graph, in $g6$ format and its color sequence, which maximizes $a(G)$ in $L_{3,n}$}.
\label{table:4-path-max}
\end{table}

\begin{table}[htb]
\scriptsize
\begin{tabular}{|p{0.03\textwidth}|p{0.06\textwidth}|p{0.5\textwidth}|p{0.28\textwidth}|}
\hline
\textbf{$n$} & \textbf{min $a(G)$} & \textbf{4-path $G$ in $g6$ format} & \textbf{$C$, color sequence for $G$} \\ \hline
10 & 2.4015 & \texttt{I\textasciitilde|xw\{N@w} & 1 2 3 4 5 \\ \hline
11 & 2.0737 & \texttt{J\textasciitilde|xw\{N@wF\_} & 1 2 3 4 5 1 \\ \hline
12 & 1.8007 & \texttt{K\textasciitilde|xw\{N@wF\_N} & 1 2 3 4 5 1 2 \\ \hline
13 & 1.5722 & \texttt{L\textasciitilde|xw\{N@wF\_N?N} & 1 2 3 4 5 1 2 3 \\ \hline
14 & 1.3804 & \texttt{M\textasciitilde|xw\{N@wF\_N?N?F\_} & 1 2 3 4 5 1 2 3 4 \\ \hline
15 & 1.2196 & \texttt{N\textasciitilde|xw\{N@wF\_N?N?F\_@w} & 1 2 3 4 5 1 2 3 4 5 \\ \hline
16 & 1.0839 & \texttt{O\textasciitilde|xw\{N@wF\_N?N?F\_@w?N} & 1 2 3 4 5 1 2 3 4 5 1 \\ \hline
17 & 0.9687 & \texttt{P\textasciitilde|xw\{N@wF\_N?N?F\_@w?N??\{} & 1 2 3 4 5 1 2 3 4 5 1 2 \\ \hline
18 & 0.8703 & \texttt{Q\textasciitilde|xw\{N@wF\_N?N?F\_@w?N??\{?@w} & 1 2 3 4 5 1 2 3 4 5 1 2 3 \\ \hline
\end{tabular}

\caption{4-path graph, in $g6$ format and its color sequence, which minimizes $a(G)$ in $L_{3,n}$}.

\label{table:4-path-min}
\end{table}

\newpage

\subsection{Maximum $\alpha$-index for $\alpha \in \{0.1, ..., 0.9\}$ of 3-paths and 4-paths of fixed order $n$}

\begin{table}[H]
  \centering
  \scriptsize
  \begin{tabular}{|c|c|c|c|c|c|c|c|c|c|l|}
    \hline
    $n$ & $A_{0.1}$ & $A_{0.2}$ & $A_{0.3}$ & $A_{0.4}$ & $A_{0.5}$ & $A_{0.6}$ & $A_{0.7}$ & $A_{0.8}$ & $A_{0.9}$ & Color sequence\\ \hline
    8  & \multicolumn{9}{|l|}{\texttt{\detokenize{G~[xhc}}}  & 1 2 1 2                          \\ \hline
    9  & \multicolumn{9}{|l|}{\texttt{\detokenize{H~[xhcp}}}  & 1 2 1 2 1                         \\ \hline
    10 & \multicolumn{9}{|l|}{\texttt{\detokenize{I~[xhcpKG}}} & 1 2 1 2 1 2                         \\ \hline
    11 & \multicolumn{9}{|l|}{\texttt{\detokenize{J~[xhcpKH__}}} & 1 2 1 2 1 2 1                       \\ \hline
    12 & \multicolumn{9}{|l|}{\texttt{\detokenize{K~[xhcpKH_e@}}} & 1 2 1 2 1 2 1 2                      \\ \hline
    13 & \multicolumn{9}{|l|}{\texttt{\detokenize{L~[xhcpKH_e@K@}}}  & 1 2 1 2 1 2 1 2 1                   \\ \hline
    14 & \multicolumn{9}{|l|}{\texttt{\detokenize{M~[xhcpKH_e@K@K?_}}} & 1 2 1 2 1 2 1 2 1 2               \\ \hline
    15 & \multicolumn{9}{|l|}{\texttt{\detokenize{N~[xhcpKH_e@K@K?e?G}}} & 1 2 1 2 1 2 1 2 1 2 1                \\ \hline
    16 & \multicolumn{9}{|l|}{\texttt{\detokenize{O~[xhcpKH_e@K@K?e?H_@}}} & 1 2 1 2 1 2 1 2 1 2 1 2              \\ \hline
    17 & \multicolumn{9}{|l|}{\texttt{\detokenize{P~[xhcpKH_e@K@K?e?H_@K?C}}} & 1 2 1 2 1 2 1 2 1 2 1 2 1            \\ \hline
    18 & \multicolumn{9}{|l|}{\texttt{\detokenize{Q~[xhcpKH_e@K@K?e?H_@K?Co?G}}} & 1 2 1 2 1 2 1 2 1 2 1 2 1 2         \\ \hline
    19 & \multicolumn{9}{|l|}{\texttt{\detokenize{R~[xhcpKH_e@K@K?e?H_@K?Co?H_?G}}} & 1 2 1 2 1 2 1 2 1 2 1 2 1 2 1      \\ \hline
  \end{tabular}
  \caption{3-paths with maximum $\alpha$-index in $L_{3,n}$. The structure of the extremal graph is the same for all $\alpha \in  \{0.1,\dots,0.9\}$.}
  \label{tab:extremal_common3}
\end{table}

\begin{table}[H]
\scriptsize
\centering
\begin{tabular}{|c|c|c|c|c|c|c|c|c|c|}
\hline
$n$ & ${0.1}$-index & ${0.2}$-index & ${0.3}$-index & ${0.4}$-index & ${0.5}$-index & ${0.6}$-index & ${0.7}$-index & ${0.8}$-index & ${0.9}$-index \\
\hline
8 & 4.8793 & 4.9308 & 4.9976 & 5.0867 & 5.2094 & 5.3832 & 5.6322 & 5.9813 & 6.4412 \\ \hline
9 & 5.1982 & 5.2751 & 5.3756 & 5.5101 & 5.6946 & 5.9504 & 6.3017 & 6.7646 & 7.3371 \\ \hline
10 & 5.4946 & 5.6007 & 5.7400 & 5.9267 & 6.1803 & 6.5245 & 6.9799 & 7.5536 & 8.2344 \\ \hline
11 & 5.7738 & 5.9124 & 6.0948 & 6.3388 & 6.6673 & 7.1037 & 7.6637 & 8.3458 & 9.1326 \\ \hline
12 & 6.0392 & 6.2130 & 6.4422 & 6.7481 & 7.1556 & 7.6868 & 8.3512 & 9.1400 & 10.0312 \\ \hline
13 & 6.2932 & 6.5046 & 6.7839 & 7.1554 & 7.6453 & 8.2729 & 9.0413 & 9.9354 & 10.9302 \\ \hline
14 & 6.5377 & 6.7888 & 7.1210 & 7.5612 & 8.1361 & 8.8612 & 9.7333 & 10.7318 & 11.8293 \\ \hline
15 & 6.7739 & 7.0667 & 7.4542 & 7.9659 & 8.6279 & 9.4512 & 10.4266 & 11.5289 & 12.7287 \\ \hline
16 & 7.0030 & 7.3391 & 7.7843 & 8.3698 & 9.1205 & 10.0427 & 11.1210 & 12.3265 & 13.6281 \\ \hline
17 & 7.2257 & 7.6068 & 8.1116 & 8.7731 & 9.6139 & 10.6353 & 11.8162 & 13.1244 & 14.5276 \\ \hline
18 & 7.4428 & 7.8703 & 8.4366 & 9.1759 & 10.1080 & 11.2288 & 12.5121 & 13.9226 & 15.4272 \\ \hline
19 & 7.6549 & 8.1300 & 8.7597 & 9.5782 & 10.6026 & 11.8231 & 13.2086 & 14.7211 & 16.3269 \\ \hline

\end{tabular}
\caption{Maximum $\alpha$-index  ($\alpha \in \{0.1,\dots,0.9\}$) for 3-paths in $L_{3,n}$, $8 \leq n \leq 19$.}
\label{tab:maximos_Q_Alpha3}
\end{table}

\begin{table}[ht]
  \centering
  \scriptsize
  \begin{tabular}{|c|c|c|c|c|c|c|c|c|c|l|}
    \hline
    $n$ & $A_{0.1}$ & $A_{0.2}$ & $A_{0.3}$ & $A_{0.4}$ & $A_{0.5}$ & $A_{0.6}$ & $A_{0.7}$ & $A_{0.8}$ & $A_{0.9}$ & Color sequence \\ \hline
    10 & \multicolumn{9}{|l|}{\texttt{\detokenize{I~|xxsxMG}}} & 1 2 1 2 1 \\ \hline
    11 & \multicolumn{9}{|l|}{\texttt{\detokenize{J~|xxsxMHo_}}} & 1 2 1 2 1 2 \\ \hline
    12 & \multicolumn{9}{|l|}{\texttt{\detokenize{K~|xxsxMHof@}}} & 1 2 1 2 1 2 1 \\ \hline
    13 & \multicolumn{9}{|l|}{\texttt{\detokenize{L~|xxsxMHof@M@}}} & 1 2 1 2 1 2 1 2\\ \hline
    14 & \multicolumn{9}{|l|}{\texttt{\detokenize{M~|xxsxMHof@M@M?_}}} & 1 2 1 2 1 2 1 2 1\\ \hline
    15 & \multicolumn{9}{|l|}{\texttt{\detokenize{N~|xxsxMHof@M@M?f?G}}} & 1 2 1 2 1 2 1 2 1 2\\ \hline
    16 & \multicolumn{9}{|l|}{\texttt{\detokenize{O~|xxsxMHof@M@M?f?Ho@}}} & 1 2 1 2 1 2 1 2 1 2 1\\ \hline
    17 & \multicolumn{9}{|l|}{\texttt{\detokenize{P~|xxsxMHof@M@M?f?Ho@M?C}}} & 1 2 1 2 1 2 1 2 1 2 1 2\\ \hline
    18 & \multicolumn{9}{|l|}{\texttt{\detokenize{Q~|xxsxMHof@M@M?f?Ho@M?Cw?G}}} & 1 2 1 2 1 2 1 2 1 2 1 2 1\\ \hline
  \end{tabular}
  \caption{4-paths with maximum $\alpha$-index in $L_{4,n}$. The structure of the extremal graph is the same for all $\alpha \in  \{0.1,\dots,0.9\}$.}
  \label{tab:extremal_common4}
\end{table}

\begin{table}[H]
\scriptsize
\centering
\begin{tabular}{|c|c|c|c|c|c|c|c|c|c|}
\hline
$n$  & ${0.1}$-index & ${0.2}$-index & ${0.3}$-index & ${0.4}$-index & ${0.5}$-index & ${0.6}$-index & ${0.7}$-index & ${0.8}$-index & ${0.9}$-index \\
\hline
10 & 6.5111 & 6.5784 & 6.6646 & 6.7775 & 6.9295 & 7.1386 & 7.4295 & 7.8290 & 8.3539 \\ \hline
11 & 6.8634 & 6.9567 & 7.0764 & 7.2333 & 7.4431 & 7.7273 & 8.1113 & 8.6167 & 9.2504 \\ \hline
12 & 7.1952 & 7.3174 & 7.4744 & 7.6799 & 7.9530 & 8.3173 & 8.7970 & 9.4076 & 10.1478 \\ \hline
13 & 7.5104 & 7.6639 & 7.8615 & 8.1198 & 8.4605 & 8.9085 & 9.4855 & 10.2008 & 11.0460 \\ \hline
14 & 7.8117 & 7.9988 & 8.2399 & 8.5543 & 8.9663 & 9.5008 & 10.1762 & 10.9953 & 11.9445 \\ \hline
15 & 8.1012 & 8.3238 & 8.6109 & 8.9847 & 9.4709 & 10.0941 & 10.8684 & 11.7910 & 12.8434 \\ \hline
16 & 8.3805 & 8.6405 & 8.9759 & 9.4115 & 9.9746 & 10.6881 & 11.5618 & 12.5873 & 13.7425 \\ \hline
17 & 8.6510 & 8.9498 & 9.3356 & 9.8356 & 10.4776 & 11.2828 & 12.2562 & 13.3843 & 14.6417 \\ \hline
18 & 8.9135 & 9.2527 & 9.6908 & 10.2572 & 10.9802 & 11.8781 & 12.9513 & 14.1817 & 15.5410 \\ \hline

\end{tabular}
\caption{Maximum $\alpha$-index ($\alpha \in \{0.1,\dots,0.9\}$) for 4-paths in $L_{4,n}$, $8 \leq n \leq 18$.}
\label{tab:maximos_Q_Alpha4}
\end{table}

\newpage

\subsection{Second maximum ${\alpha}$-index for $\alpha \in \{0.1, ..., 0.9\}$ of 3-paths and 4-paths of fixed order $n$}

\begin{table}[H]
  \centering
  \scriptsize
  \begin{tabular}{|c|c|c|c|c|c|c|c|c|c|l|}
    \hline
    $n$  & $A_{0.1}$ & $A_{0.2}$ & $A_{0.3}$ & $A_{0.4}$ & $A_{0.5}$ & $A_{0.6}$ & $A_{0.7}$ & $A_{0.8}$ & $A_{0.9}$ & Color sequence \\ \hline
    8 & \multicolumn{9}{|l|}{\texttt{\detokenize{G~[xgk}}} & 1 2 1 3 \\ \hline
    9 & \multicolumn{9}{|l|}{\texttt{\detokenize{H~[xhcR}}} & 1 2 1 2 3 \\ \hline
    10 & \multicolumn{9}{|l|}{\texttt{\detokenize{I~[xhcpCW}}} & 1 2 1 2 1 3 \\ \hline
    11 & \multicolumn{9}{|l|}{\texttt{\detokenize{J~[xhcpKG`_}}} & 1 2 1 2 1 2 3 \\ \hline
    12 & \multicolumn{9}{|l|}{\texttt{\detokenize{K~[xhcpKH_aB}}} & 1 2 1 2 1 2 1 3 \\ \hline
    13 & \multicolumn{9}{|l|}{\texttt{\detokenize{L~[xhcpKH_e@CB}}} & 1 2 1 2 1 2 1 2 3 \\ \hline
    14 & \multicolumn{9}{|l|}{\texttt{\detokenize{M~[xhcpKH_e@K@C@_}}} & 1 2 1 2 1 2 1 2 1 3 \\ \hline
    15 & \multicolumn{9}{|l|}{\texttt{\detokenize{N~[xhcpKH_e@K@K?a?W}}} & 1 2 1 2 1 2 1 2 1 2 3 \\ \hline
    16 & \multicolumn{9}{|l|}{\texttt{\detokenize{O~[xhcpKH_e@K@K?e?G_B}}} & 1 2 1 2 1 2 1 2 1 2 1 3 \\ \hline
    17 & \multicolumn{9}{|l|}{\texttt{\detokenize{P~[xhcpKH_e@K@K?e?H_@C?K}}} & 1 2 1 2 1 2 1 2 1 2 1 2 3 \\ \hline
    18 & \multicolumn{9}{|l|}{\texttt{\detokenize{Q~[xhcpKH_e@K@K?e?H_@K?CO?W}}} & 1 2 1 2 1 2 1 2 1 2 1 2 1 3 \\ \hline
    19 & \multicolumn{9}{|l|}{\texttt{\detokenize{R~[xhcpKH_e@K@K?e?H_@K?Co?G_?W}}} & 1 2 1 2 1 2 1 2 1 2 1 2 1 2 3 \\ \hline
  \end{tabular}
  \caption{3-paths with maximum  ${\alpha}$-index in $L_{3,n} \backslash \{K_{2} \vee P_{n-2}\}$. The structure of the extremal graph is the same for all $\alpha \in  \{0.1,\dots,0.9\}$.}
  \label{tab:extremal_common3alpha_2}
\end{table}

\begin{table}[H]
\scriptsize
\centering
\begin{tabular}{|c|c|c|c|c|c|c|c|c|c|}
\hline
$n$ &  {second} &  {second-} &  {second-} &  {second-} &  {second-} &  {second-} &  {second-} &  {second-} &  {second-} \\
 &  {max} &  {max} &  {max} &  {max} &  {max} &  {max} &  {max} &  {max} &  {max} \\
 &  {${0.1}$-index} &  {$0.2$-index} &  {$0.3$-index} &  {${0.4}$-index} &  {${0.5}$-index} &  {${0.6}$-index} &  {${0.7}$-index} &  {${0.8}$-index} &  {${0.9}$-index} \\
\hline
8 & 4.8208 & 4.8612 & 4.9132 & 4.9825 & 5.0791 & 5.2216 & 5.4452 & 5.8063 & 6.3397 \\ \hline
9  & 5.1250 & 5.1877 & 5.2694 & 5.3796 & 5.5337 & 5.7579 & 6.0923 & 6.5823 & 7.2354 \\ \hline
10  & 5.4123 & 5.5020 & 5.6198 & 5.7791 & 6.0006 & 6.3148 & 6.7603 & 7.3691 & 8.1331 \\ \hline
11  & 5.6857 & 5.8061 & 5.9650 & 6.1798 & 6.4753 & 6.8838 & 7.4389 & 8.1607 & 9.0316 \\ \hline
12  & 5.9474 & 6.1016 & 6.3058 & 6.5810 & 6.9554 & 7.4606 & 8.1237 & 8.9549 & 9.9306 \\ \hline
13  & 6.1991 & 6.3897 & 6.6426 & 6.9824 & 7.4392 & 8.0424 & 8.8123 & 9.7506 & 10.8298 \\ \hline
14  & 6.4420 & 6.6714 & 6.9761 & 7.3839 & 7.9258 & 8.6279 & 9.5034 & 10.5473 & 11.7292 \\ \hline
15  & 6.6773 & 6.9475 & 7.3066 & 7.7853 & 8.4145 & 9.2161 & 10.1963 & 11.3446 & 12.6288 \\ \hline
16 & 6.9058 & 7.2186 & 7.6346 & 8.1867 & 8.9049 & 9.8062 & 10.8904 & 12.1425 & 13.5284 \\ \hline
17  & 7.1283 & 7.4854 & 7.9604 & 8.5880 & 9.3965 & 10.3978 & 11.5856 & 12.9407 & 14.4281 \\ \hline
18  & 7.3454 & 7.7483 & 8.2842 & 8.9891 & 9.8892 & 10.9907 & 12.2815 & 13.7393 & 15.3278 \\ \hline
19  & 7.5575 & 8.0076 & 8.6062 & 9.3902 & 10.3828 & 11.5844 & 12.9780 & 14.5380 & 16.2276 \\ \hline

\end{tabular}
\caption{Maximum ${\alpha}$-index ($\alpha \in \{0.1,\dots,0.9\}$) for 3-paths in $L_{3,n} \backslash \{K_{2} \vee P_{n-2}\}$, $8 \leq n \leq 19$.}
\label{tab:maximos_Q_Alpha3alpha_2}
\end{table}

\begin{table}[H]
  \centering
  \scriptsize
  \begin{tabular}{|c|c|c|c|c|c|c|c|c|c|l|}
    \hline
    $n$  & $A_{0.1}$ & $A_{0.2}$ & $A_{0.3}$ & $A_{0.4}$ & $A_{0.5}$ & $A_{0.6}$ & $A_{0.7}$ & $A_{0.8}$ & $A_{0.9}$ & Color sequence\\ \hline
    10 & \multicolumn{9}{|l|}{\texttt{\detokenize{I~|xxsxEW}}} & 1 2 1 2 3 \\ \hline
    11 & \multicolumn{9}{|l|}{\texttt{\detokenize{J~|xxsxMGp_}}} & 1 2 1 2 1 3 \\ \hline
    12 & \multicolumn{9}{|l|}{\texttt{\detokenize{K~|xxsxMHobB}}} & 1 2 1 2 1 2 3 \\ \hline
    13 & \multicolumn{9}{|l|}{\texttt{\detokenize{L~|xxsxMHof@EB}}} & 1 2 1 2 1 2 1 3 \\ \hline
    14 & \multicolumn{9}{|l|}{\texttt{\detokenize{M~|xxsxMHof@M@E@_}}} & 1 2 1 2 1 2 1 2 3 \\ \hline
    15 & \multicolumn{9}{|l|}{\texttt{\detokenize{N~|xxsxMHof@M@M?b?W}}} & 1 2 1 2 1 2 1 2 1 3 \\ \hline
    16 & \multicolumn{9}{|l|}{\texttt{\detokenize{O~|xxsxMHof@M@M?f?GoB}}} & 1 2 1 2 1 2 1 2 1 2 3 \\ \hline
    17 & \multicolumn{9}{|l|}{\texttt{\detokenize{P~|xxsxMHof@M@M?f?Ho@E?K}}} & 1 2 1 2 1 2 1 2 1 2 1 3 \\ \hline
    18 & \multicolumn{9}{|l|}{\texttt{\detokenize{Q~|xxsxMHof@M@M?f?Ho@M?CW?W}}} & 1 2 1 2 1 2 1 2 1 2 1 2 3 \\ \hline
  \end{tabular}
  \caption{4-path with maximum ${\alpha}$-index in $L_{4,n} \backslash \{K_{3} \vee P_{n-3}\}$. The structure of the extremal graph is the same for all $\alpha \in  \{0.1,\dots,0.9\}$. }
  \label{tab:extremal_common4alpha}
\end{table}

\begin{table}[H]
\scriptsize
\centering
\begin{tabular}{|c|c|c|c|c|c|c|c|c|c|}
\hline
$n$ &  {second} &  {second-} &  {second-} &  {second-} &  {second-} &  {second-} &  {second-} &  {second-} &  {second-} \\
 &  {max} &  {max} &  {max} &  {max} &  {max} &  {max} &  {max} &  {max} &  {max} \\
 &  {${0.1}$-index} &  {$0.2$-index} &  {$0.3$-index} &  {${0.4}$-index} &  {${0.5}$-index} &  {${0.6}$-index} &  {${0.7}$-index} &  {${0.8}$-index} &  {${0.9}$-index} \\
\hline
10  & 6.4563 & 6.5149 & 6.5898 & 6.6885 & 6.8228 & 7.0124 & 7.2880 & 7.6928 & 8.2652 \\ \hline
11  & 6.8015 & 6.8847 & 6.9915 & 7.1325 & 7.3234 & 7.5881 & 7.9599 & 8.4762 & 9.1612 \\ \hline
12  & 7.1286 & 7.2396 & 7.3826 & 7.5711 & 7.8245 & 8.1699 & 8.6399 & 9.2651 & 10.0587 \\ \hline
13  & 7.4406 & 7.5821 & 7.7648 & 8.0051 & 8.3258 & 8.7556 & 9.3250 & 10.0572 & 10.9569 \\ \hline
14  & 7.7397 & 7.9142 & 8.1396 & 8.4355 & 8.8272 & 9.3442 & 10.0135 & 10.8512 & 11.8555 \\ \hline
15  & 8.0278 & 8.2372 & 8.5080 & 8.8626 & 9.3285 & 9.9348 & 10.7042 & 11.6465 & 12.7545 \\ \hline
16  & 8.3061 & 8.5523 & 8.8710 & 9.2871 & 9.8297 & 10.5269 & 11.3967 & 12.4428 & 13.6536 \\ \hline
17  & 8.5759 & 8.8606 & 9.2292 & 9.7092 & 10.3309 & 11.1202 & 12.0904 & 13.2397 & 14.5529 \\ \hline
18  & 8.8381 & 9.1627 & 9.5832 & 10.1294 & 10.8319 & 11.7144 & 12.7851 & 14.0371 & 15.4524 \\ \hline

\end{tabular}
\caption{Maximum ${\alpha}$-index ($\alpha \in \{0.1,\dots,0.9\}$) for 4-paths in $L_{4,n} \backslash \{K_{3} \vee P_{n-3}\}$, $10 \leq n \leq 18$.}
\label{tab:maximos_Q_Alpha4alpha}
\end{table}

\end{document}